\documentclass[sigconf]{acmart}

\usepackage[ruled,noend]{algorithm2e}
\usepackage{booktabs} 
\usepackage{multirow}
\usepackage{array}
\usepackage{amsmath,amssymb,amsfonts}
\usepackage{algorithmic}
\usepackage{graphicx}
\usepackage{textcomp}
\usepackage{soul}
\usepackage[labelformat=simple]{subcaption}

\ifodd 1
\usepackage{soul, xcolor}

\newcommand{\rev}[1]{{{\color{black} #1}}}

\fi

\newcommand{\shaun}[1]{\textcolor{black}{#1}}

\setcopyright{rightsretained}

\acmDOI{10.475/123_4}

\acmISBN{123-4567-24-567/08/06}

\acmConference[ICCPS'19]{ACM ICCPS}{April 2019}{Montreal, Canada}
\acmYear{2019}
\copyrightyear{2019}

\acmArticle{4}
\acmPrice{15.00}

\begin{document}
\title{TACAN: Transmitter Authentication through Covert Channels in Controller Area Networks}

\author{Xuhang Ying}
\affiliation{%
  \institution{Department of Electrical and Computer Engineering \\ University of Washington}
  \city{Seattle}
  \state{WA 98195}
}
\email{xhying@uw.edu}

\author{Giuseppe Bernieri}
\affiliation{%
  \institution{Department of Mathematics \\ University of Padua}
  \city{Padua}
  \state{Italy}
}
\email{bernieri@math.unipd.it}

\author{Mauro Conti}
\affiliation{%
  \institution{Department of Mathematics \\ University of Padua}
  \city{Padua}
  \country{Italy}}
\email{conti@math.unipd.it}


\author{Radha Poovendran}
\affiliation{%
  \institution{Department of Electrical and Computer Engineering \\ University of Washington}
  \city{Seattle}
  \state{WA 98195}
}
\email{rp3@uw.edu}

\renewcommand{\shortauthors}{X. Ying et al.}

\begin{abstract}
Nowadays, the interconnection of automotive systems with modern digital devices offers advanced user experiences to drivers.
Electronic Control Units (ECUs) carry out a multitude of operations using the insecure Controller Area Network (CAN) bus in automotive Cyber-Physical Systems (CPSs). Therefore, dangerous attacks, such as disabling brakes, are possible and the safety of passengers is at risk.
In this paper, we present TACAN (Transmitter Authentication in CAN), which provides secure authentication of ECUs by exploiting the \textit{covert channels} without introducing CAN protocol modifications or traffic overheads \rev{(i.e., no extra bits or messages are used)}. 
TACAN turns upside-down the originally malicious concept of covert channels and exploits it to build an effective defensive technique that facilitates transmitter authentication via a trusted Monitor Node.
TACAN consists of three different covert channels for ECU authentication:
1) Inter-Arrival Time (IAT)-based, leveraging the IATs of CAN messages; 2) offset-based, exploiting the clock offsets of CAN messages; 3) Least Significant Bit (LSB)-based, concealing authentication messages into the LSBs of normal CAN data.
We implement the covert channels of TACAN on the University of Washington (UW) EcoCAR testbed and evaluate their performance through extensive experiments using real vehicle datasets. 
We demonstrate the \rev{feasibility} 
of TACAN, highlighting no traffic overheads and attesting the regular functionality of ECUs.
In particular, the bit error \rev{ratios} are within $0.1\%$ and $0.42\%$ for the IAT-based and offset-based covert channels, respectively. 
Furthermore, the bit error \rev{ratio} of the LSB-based covert channel is equal to that of a normal CAN bus, which is $3.1\times 10^{-7}$\%.
\end{abstract}

%
%



\maketitle

\section{Introduction}
The Controller Area Network (CAN) enables communication among Electronic Control Units (ECUs) for closed in-vehicle networks~\cite{ISO2015CAN,bosch1991CAN}. 
The security of CAN bus is crucial to the functionality and safety of today's automobiles and future's autonomous cars~\cite{bojarski2016end,wyglinski2013security}.
Since the CAN bus is a broadcast medium without authentication, a compromised ECU can be used to masquerade as any targeted ECU by transmitting messages with the forged message ID (masquerade attack~\cite{checkoway2011comprehensive}).
Modern externally accessible ECUs
\rev{with additional connectivity interfaces such as cellular, Wi-Fi or Bluetooth} 
disrupt the closed in-vehicle network assumption. Consequently, the CAN bus \shaun{has shown to be} vulnerable to cyber attacks, such as disabled brakes~\cite{checkoway2011comprehensive} and remotely controlled steering~\cite{tesla2016remote}.

\rev{Use of cryptographic primitives such as message authentication is one way to defend against attacks (notably the masquerade attack) on the CAN bus. 
However, it can be challenging in practice} due to the low throughput and tight bit budget of the CAN protocol, 
and current solutions such as~\cite{herrewege2011canauth,hazem2012lcap,kurachi2014cacan,radu2016leia}  \rev{require protocol modifications or introduce traffic overheads.}
An alternative is to deploy anomaly-based Intrusion Detection Systems (IDSs) without modifying the CAN protocol~\cite{cho2016finger,muter2011entropy,cho2017viden,choi2018voltageids}, \rev{including} timing-based and voltage-based IDSs.
\shaun{The timing-based IDS} in~\cite{cho2016finger} exploits CAN message periodicity to estimate clock skew as a unique fingerprint to detect masquerade attacks. 
Nevertheless, it was later shown to be ineffective against \rev{the cloaking attack} that modifies the inter-transmission time to emulate the clock skew of the targeted ECU~\cite{sagong2018cloaking,ying2018shape}. 
The voltage-based IDSs~\cite{murvay2014source,cho2017viden,choi2018voltageids,kneib2018scission} \rev{attempt to} 
fingerprint the attacker through voltage signal characteristics. 
\rev{However, if the attacker uses IDs that the compromised ECU is allowed to use under normal conditions, the attack will not be detected.}


\begin{figure}[t!]
\centering
\includegraphics[width=1\columnwidth]{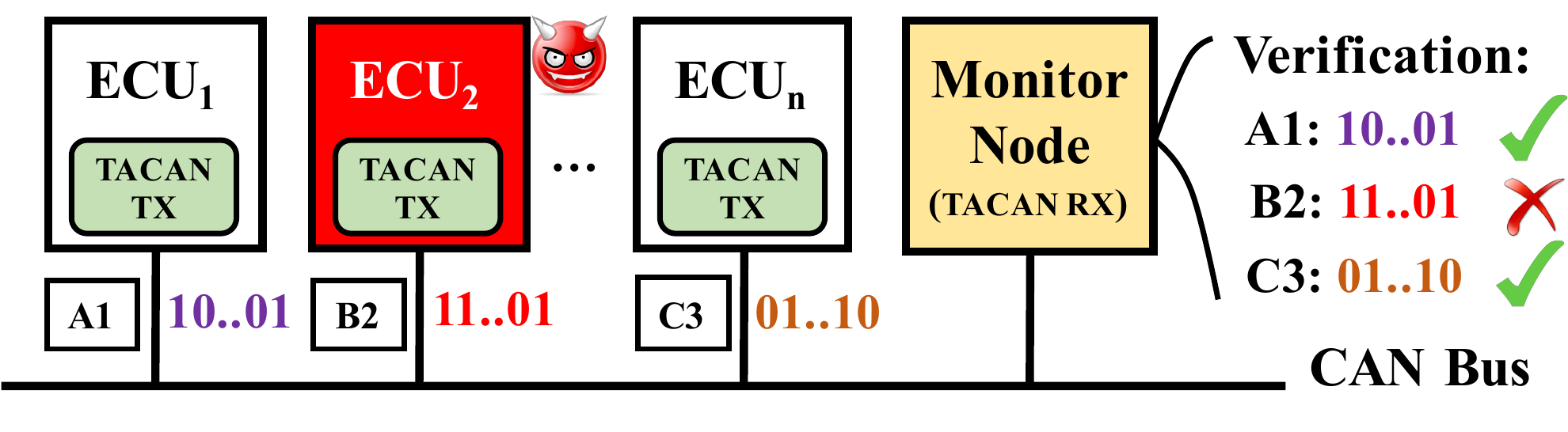}
\caption{Illustration of TACAN. Legitimate ECUs transmit unique authentication messages that are embedded into the \rev{timing and LSBs} of normal CAN messages \rev{(e.g., A1, B2, and C3)} using the proposed covert channel methodologies. 
The Monitor Node authenticates transmitting ECUs by verifying the received authentication messages. \rev{If the} compromised ECUs cannot generate valid authentication messages, then the attack will be detected by the Monitor Node.}
\label{fig:main_idea}
\end{figure}

In this work, we develop TACAN that allows the trusted Monitor Node (MN) to verify the authenticity of a transmitting ECU \rev{and} detect CAN bus anomalies, as illustrated in Figure~\ref{fig:main_idea}.
In TACAN, a master key is shared between an ECU and the MN for generating shared session keys. 
\rev{Consistently with \cite{hazem2012lcap,radu2016leia, herrewege2011canauth}, we assume that the keys are stored in the tamper-resistant memory of a security module such as the Trusted Platform Module (TPM) \cite{TPMautomotive}. 
}
%
%
%
%
Each ECU embeds unique authentication messages into CAN messages and \rev{continuously} transmits them through covert channels, which can be received and verified by the MN.

\rev{Therefore, if} the attacker has no access to the TPM \rev{of the targeted ECU}, 
it cannot use the compromised ECU or external device to generate valid authentication 
\rev{messages}, thus causing verification failures and triggering the alarm at the MN side.
\rev{Moreover, TACAN will detect attacks (e.g., \rev{Denial-of-Service (DoS)} attack) that interrupt the transmission of CAN messages that are used to embed the authentication messages.}
\rev{The main benefits of using covert channels for TACAN are that they do not introduce protocol modifications or traffic overheads (i.e., extra bits or messages).
In addition, by requiring ECUs to transmit authentication messages much less frequently than per-message authentication schemes, TACAN can significantly reduce the computational burden of the resource-constrained ECUs.}


\vspace{0.1cm}
\noindent \textbf{Contributions: }
In this paper, we make the following contributions: 
\begin{itemize}
    \item \shaun{We identify and exploit covert channels to facilitate ECU authentication on the CAN bus. 
    Hence, covert channels are used for security instead of malicious communication.}
    
    
    \item We propose \shaun{TACAN, which consists of a suite of} three covert channels for transmitting authentication messages, including two timing-based covert channels that modify the inter-transmission times of CAN messages to affect the inter-arrival times (IATs) and offsets observed by the MN, and one storage-based covert channel that hides the information in the LSBs of the data payload of normal CAN messages.

    \item We implement the covert channels of TACAN in a real vehicle testbed (the UW EcoCAR~\cite{ecocar}).
    We also conduct extensive experiments 
    \rev{to demonstrate the feasibility of such covert channels}
    using two real CAN traffic datasets, the publicly available Toyota dataset~\cite{toyota2010dataset} and the EcoCAR dataset. 
    Our results show that the bit error \rev{ratios} are within $0.1\%$ and $0.42\%$ for the IAT-based and offset-based covert channels, respectively.
    The bit error \rev{ratio} of the LSB-based covert channel is equal to $3.1\times 10^{-7}$\%, which is the bit error \rev{ratio} of a typical CAN bus.
\end{itemize}

\noindent \textbf{Organization.} The remainder of this paper is organized as follows.
Section~\ref{sec:related_work} reviews the related work.
Section~\ref{sec:system_and_adversary_model} presents our system and adversary models.
Section~\ref{sec:proposed_scheme} presents TACAN. 
A security discussion of TACAN is provided in Section~\ref{sec:security_analysis}. 
Section~\ref{sec:evaluation} presents experimental evaluation.
Section~\ref{sec:conclusion} concludes this paper. 

\section{Related Work}\label{sec:related_work}
Recent experimental studies have demonstrated that an attacker is able to infiltrate in-vehicle ECUs physically or remotely and mount cyber attacks that would cause potentially life-threatening consequences by disabling breaks or overriding steering~\cite{miller2015remote,checkoway2011comprehensive}.
\shaun{One way to secure the CAN bus is to deploy anomaly-based IDSs} 
based on traffic analysis (e.g., timing/frequency~\cite{hoppe2008security}), 
entropy~\cite{muter2011entropy}, or physical invariants such as clock skew~\cite{cho2016finger,sagong2018cloaking} and signal characteristics~\cite{murvay2014source,cho2017viden,choi2018voltageids,kneib2018scission}. 
While voltage-based IDSs are effective against ongoing masquerade attackers, they cannot detect a compromised ECU before attacks are launched (e.g., a stealthy attacker may not launch the attack until the car is in drive mode). 
\rev{Moreover, it has been recently shown in \cite{sagong2018exploring} that the extra wires required by voltage-based IDSs may introduce new attack surfaces.}

Researchers have also attempted to add cryptographic primitives such as Message Authentication Code (MAC) to the CAN bus, including CANAuth~\cite{herrewege2011canauth}, LCAP~\cite{hazem2012lcap}, CaCAN~\cite{kurachi2014cacan}, and LeiA~\cite{radu2016leia}. 
Due to the tight bit budget and low throughput of CAN, authentication information is usually embedded into existing CAN messages (i.e., the ID or data field) and transmitted through additional CAN messages, thus introducing \rev{traffic}
overheads or increasing the bus load~\cite{hazem2012lcap,kurachi2014cacan,radu2016leia}.
\shaun{In this work, we focus on transmitter authentication rather than per-message authentication \rev{to avoid such traffic overheads}. 
The key novelty of this work is the use of covert channels, a well-known malicious technique that is converted into defensive applications for authentication purposes.
Compared to previous authentication schemes, our scheme does not require protocol modifications or introduce extra bits or CAN messages.}


\shaun{In literature, there are two main categories of covert channels: timing-based and storage-based. 
In timing-based covert channels, only the timing of events or traffic is modified to share information between two parties and the contents of data stream remain intact.
The storage-based covert channels hide data in a shared resource that is not designed for transferring data, e.g., by exploiting reserved or used fields in the data packet or concealing data in the payload.
Compared to steganography techniques that require some form of content as cover, the covert channels exploit network protocols as carrier~\cite{zander2007survey}.
In~\cite{taylor2017enhancing}, Taylor \textit{et al.} propose an approach where the exploitation of covert channel enhances the Modbus/TCP protocol security for industrial control system applications.
To the best of our knowledge, this is the first paper that explores covert channels for automotive CAN buses. 
}

\section{System and Adversary Models}
\label{sec:system_and_adversary_model}
In this section, we present the system (Section~\ref{sec:system_model}) and adversary (Section~\ref{sec:adversary_model}) models for the CAN bus.
A list of frequently used notations is provided in Table~\ref{table:notation}.

\begin{table}[t!]
	\footnotesize
	\centering
	\caption{Frequently used notations.}
	\vspace{-0.1cm}
	\begin{tabular}{|c|l|}
		\hline
		\textbf{Notation} & \textbf{Description} \\
		\hline
		$MK_i, SK_i$ & Master and session keys of $ECU_i$ \\ \hline
		$g_i, l_i$ & Global and local counters of $ECU_i$ \\ \hline
		$T$ & Message period\\ \hline
		$S$ & Clock skew \\ \hline
		$t_i, a_i$ & Transmit and arrival timestamps \\ \hline
		$\eta_i$ & Noise in the arrival timesatmp of message $i$ \\ \hline
		$\Delta t_i$ & Inter-transmission time (ITT) between messages $(i-1)$ and $i$\\ \hline
		$\Delta a_i$ & Inter-arrival time (IAT) between messages $(i-1)$ and $i$ \\ \hline
		$\Delta \bar{a}[i]$ & The $i$-th sample of averaged $\Delta a_i$\\ \hline 
		$\delta$ & Deviation (added to ITTs at the transmitter side) \\ \hline
		$L$ & Window length or number of least significant bits \\ \hline
		$\kappa, \Gamma_u, \Gamma_l$ & Reference level, upper threshold, lower threshold \\ \hline
		$A_m$, $A_f$ & Authentication message, authentication frame \\ \hline
		$n_m$, $n_f$, $n_s$ & Number of bits in $A_m$ and $A_f$; number of silence bits\\ \hline 
		$\hat{O}_k[i]$ & The observed clock offset up to message $i$ in batch $k$\\ \hline 
		
	\end{tabular}
	
	\label{table:notation}
	\normalsize
\end{table}

\subsection{System Model}\label{sec:system_model}

\noindent \textbf{CAN Bus.}
As a broadcast medium, the CAN bus connects all ECUs to the same, shared bus line and allows them to transmit any messages to any ECU and observe all ongoing transmissions. 
Each CAN frame (or message) has a set of predefined fields, including notably the Arbitration field (which includes the message ID) 
and the Data field (up to 8 bytes).
More details about the CAN frame structure are provided in Appendix~\ref{appendix:CAN_frame}. 

If two (or more) ECUs attempt to transmit messages simultaneously, an arbitration scheme based on priority (a smaller message ID indicates a higher priority) is used to determine which ECU transmits first. 
The CAN messages do not have transmit timestamps and do not support encryption or authentication. 

\vspace{0.1cm}
\noindent \textbf{Clock Skew.} 
In automotive CAN, \rev{most} messages are transmitted periodically as per the local clocks of transmitting ECUs\footnote{\rev{In the UW EcoCAR (Chevrolet Camaro), all of the 89 messages with distinct IDs are periodic with periods ranging from 10 ms to 5 sec. In the Toyota Camry 2010, 39 of the 42 distinct messages are periodic. In the Dodge Ram Pickup 2010 in \cite{cho2016finger}, all of the 55 distinct messages are periodic. While CAN message periodicity depends on the manufacturer and the model, the above examples suggest that periodic CAN messages are very common and even dominant on the CAN bus of commercial automobiles.}}. 
Since there exists no clock synchronization among ECUs, the frequencies of local clocks are different due to clock skew, a physical property caused by variations in the clock's hardware crystal. 

Let $C_A(t)$ be the time reported by clock $A$ and $C_{true}(t)=t$ be the true time. 
According to the Network Time Protocol (NTP)~\cite{mills1992NTP}, the \textit{clock offset} of clock A is defined as $O_A(t)=C_A(t)-C_{true}(t)$, 
and the \textit{clock skew} is the first derivative of clock offset, i.e., $S_A(t)=O'_A(t)=C'_A(t)-1$, which is usually measured in microseconds per second ($\mu$s/s) or parts per million (ppm).
In the absence of a true clock, the \textit{relative clock offset} and \textit{relative clock skew} can be defined with respect to a reference clock. 

\begin{figure}[t!]
\centering
\includegraphics[trim=0cm 0.9cm 0cm 0cm, clip=true, width=.9\columnwidth]{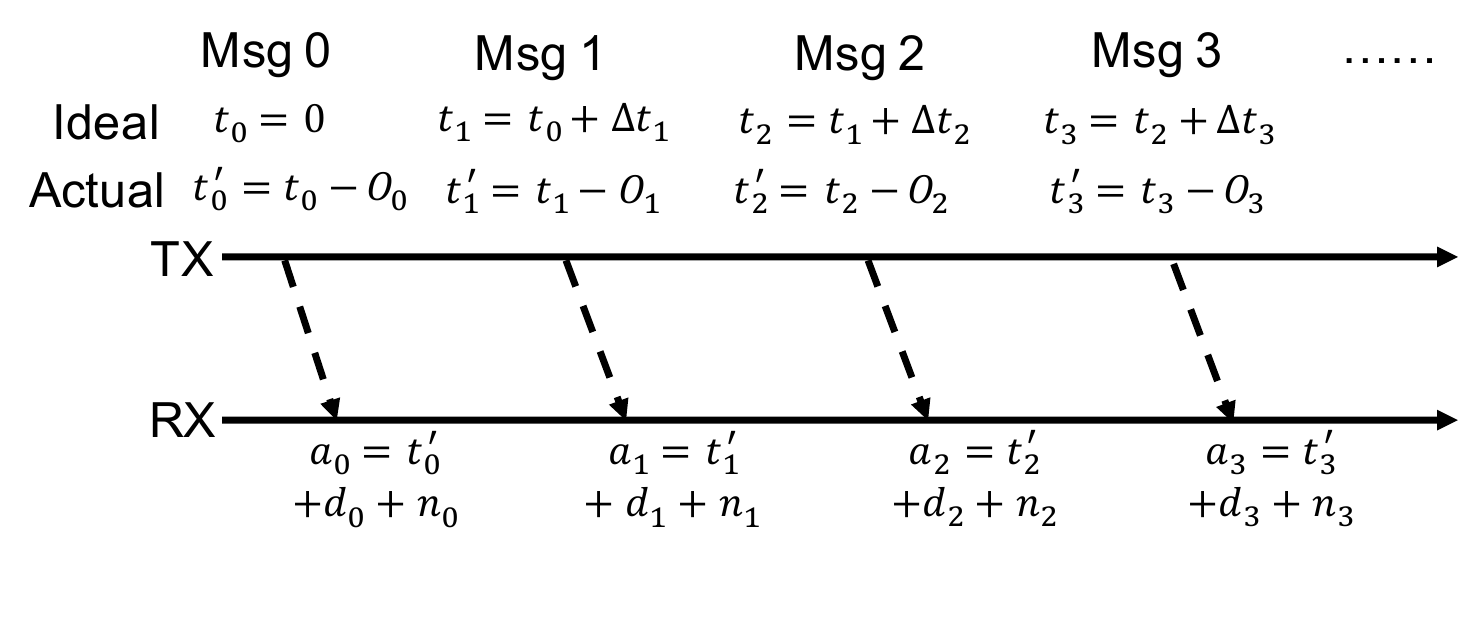}
\caption{Illustration of our timing model. 
}
\label{fig:timing_model}
\vspace{-0.3cm}
\end{figure}

\vspace{0.1cm}
\noindent \textbf{Timing Model.}
As illustrated in Figure~\ref{fig:timing_model}, we let $t_i$ be the transmit time of message $i$ (assuming $t_0=0$) and $\Delta t_i = t_i - t_{i-1}$ be the inter-transmission time according to the transmitter's clock.
If messages are transmitted every $T$ seconds, we have $\Delta t_i=T$ and $t_i=iT$. 
The \rev{receiver's clock} is considered as the reference clock. 
In practice, there exists a clock skew in the transmitter's clock relative to the reference clock, which introduces an offset $O_i$ between the two clocks. 
Therefore, the actual transmit time is $t'_i = t_i - O_i$ \rev{according to the reference clock}. 

\rev{While the clock skew may be slowly varying due to factors including the temperature, the clock skew is almost constant over a short duration.}
Given a constant clock skew  
$S$, the relationship between the elapsed time $t_i$ at the transmitter and the elapsed time $t'_i$ at the receiver is given by $S = (t_i - t'_i)/t'_i$.
Hence, we have $t_i' = t_i/(1+S)$, and $O_i = t_i - t_i' = \frac{S}{1+S} t_i$.
To account for offset deviations due to jitters, we model the actual clock offset $O_i = \frac{S}{1+S} t_i + \epsilon_i$, 
where $\epsilon_i$'s are assumed to be i.i.d. zero-mean random variables.

After a network delay of $d_i$ (due to message transmission, propagation, arbitration, and reception) and the zero-mean quantization noise $n_i$~\cite{zander2008measurement}, the arrival timestamp of message $i$ is
\begin{align}
    a_i = t_i - O_i + d_i + n_i &=t_i - \frac{S}{1+S} t_i  + \eta_i =\frac{1}{1+S}t_i + \eta_i, \label{eq:arrival_timestamp}
\end{align}
where $\eta_i = -\epsilon_i + d_i + n_i$ is the overall noise. 
Since periodic CAN messages have constant data lengths over time, it is reasonable to assume constant-mean network delays, i.e., $\mathbb{E}[d_i] = d$. 
Hence, $\eta_i$'s can be modeled as i.i.d. random variables with a mean of $d$ and a variance of $\sigma_\eta^2$.

\subsection{Adversary Model}
\label{sec:adversary_model}
We consider an adversary \shaun{who aims to infiltrate the CAN bus and launch stealthy attacks without being detected.}
We assume that the adversary can passively monitor and observe all ongoing CAN transmissions. 
In addition, it has full knowledge of the deployed covert channels and can also observe all authentication messages that are being transmitted.
In reality, there are usually two ways of gaining unauthorized access to the CAN bus: 1) compromise an in-vehicle ECU either physically or remotely~\cite{checkoway2011comprehensive}, or 2) plug an external device (a malicious ECU) into the CAN bus~\cite{koscher2010experimental}.
\rev{We assume that the adversary has no access to the keys stored in the TPM of the compromised ECU and other legitimate ECUs.}

\begin{figure}[t!]
    \centering
    \begin{subfigure}[b]{.85\columnwidth}
        \includegraphics[width=\textwidth]{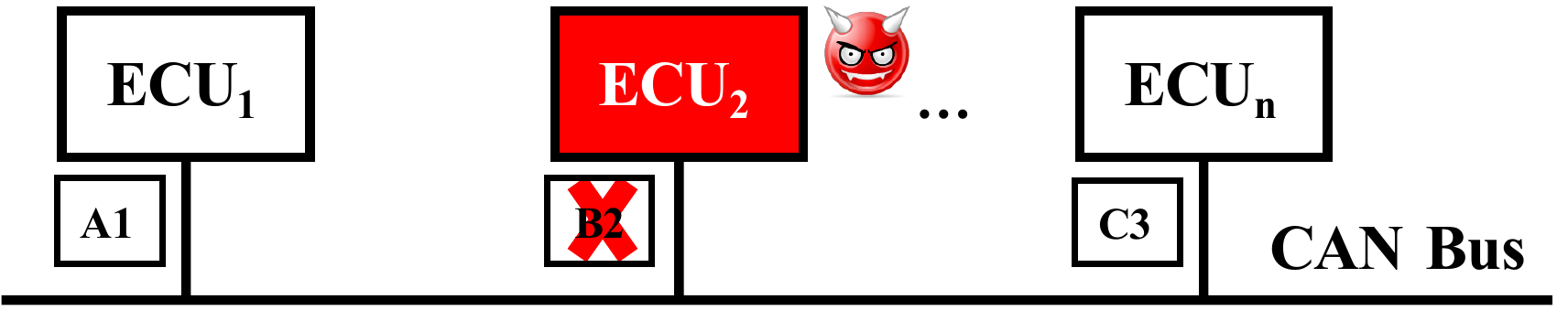}
        \vspace{-0.5cm}
        \caption{Suspension attack}
        \label{fig:attack_suspension}
    \end{subfigure}
    
    \vspace{0.2cm}
    
    \begin{subfigure}[b]{.85\columnwidth}
        \includegraphics[width=\textwidth]{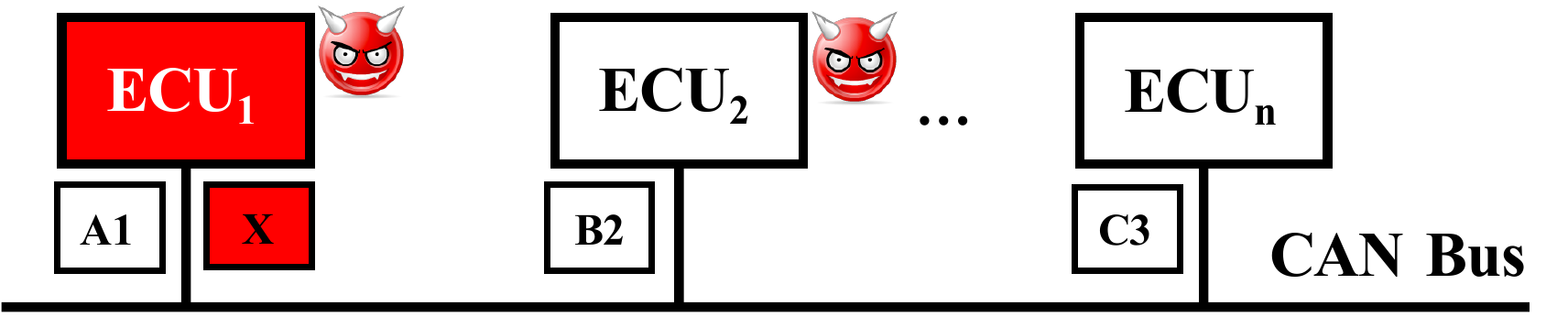}
        \vspace{-0.5cm}
        \caption{Injection attack}
        \label{fig:attack_fabrication}
    \end{subfigure}
    
    \vspace{0.2cm}
    
    \begin{subfigure}[b]{.85\columnwidth}
        \includegraphics[width=\textwidth]{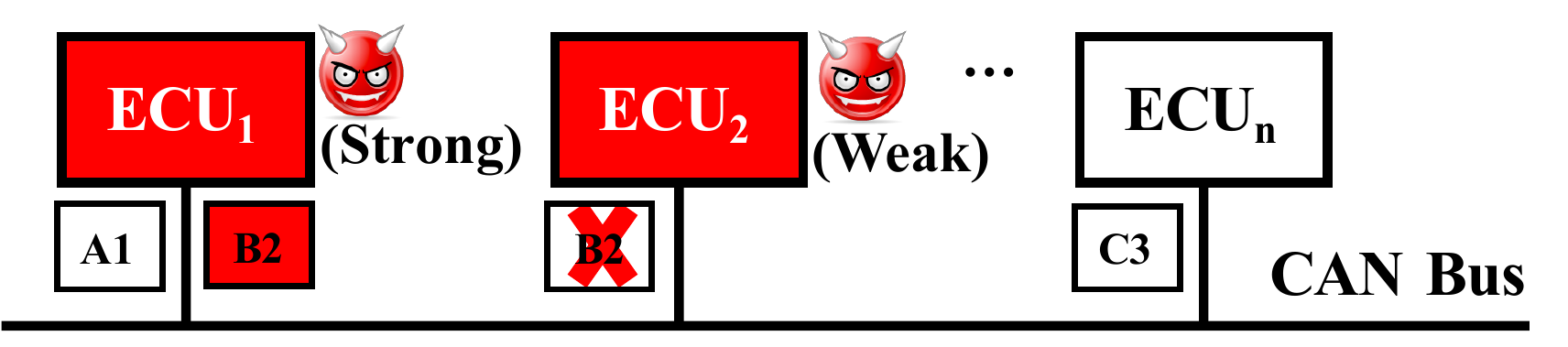}
        \vspace{-0.5cm}
        \caption{Masquerade attack}
        \label{fig:attack_masquerade}
    \end{subfigure}
    
    \vspace{-0.2cm}
    \caption{Three representative attacks on the CAN bus. 
    }
    \label{fig:attack_model}
\end{figure}

The adversary can use the compromised or malicious ECU to perform three basic attacks: 1) suspension, 2) injection, and 3) masquerade attacks, as considered in~\cite{lin2012cyber,cho2016finger,sagong2018cloaking}.
As illustrated in Figure~\ref{fig:attack_suspension}, a suspension attacker prevents the compromised ECU from transmitting certain messages, while an injection attacker can fabricate and inject CAN messages of arbitrary choices of message ID, content, and timing, as sketched in Figure~\ref{fig:attack_fabrication}.
Injection attacks can lead to more sophisticated attacks such as the \rev{DoS} 
attack~\cite{hoppe2008security} and the bus-off attack~\cite{cho2016error}. 
In the (stealthy) masquerade attack, the adversary will need to compromise two ECUs -- one is weakly compromised (acting as the strong attacker who can only launch suspension attacks) and the other one is fully compromised (acting as the strong attacker who can both launch suspension and injection attacks). 
In the example in Figure~\ref{fig:attack_masquerade}, the adversary suspends the weakly compromised $ECU_2$ from transmitting message 0xB2 and uses the fully compromised $ECU_1$ to inject messages 0xB2 claiming to originate from $ECU_2$. 
As compared to the suspension and injection attack, it is much more difficult to detect the stealthy masquerade attack.

\section{TACAN}
\label{sec:proposed_scheme}
In this section, we provide details on the TACAN architecture (Section~\ref{sec:tacan_architecture}) and introduce our transmitter authentication protocol (Section~\ref{sec:transmission_authentication_protocol}).
We then present three covert channels for transmitting authentication messages: 
1) IAT-based (Section~\ref{sec:IAT_based_covert_channel}), 2) offset-based (Section~\ref{sec:offset_based}), and 3) LSB-based (Section~\ref{sec:lsb_based}).

\subsection{TACAN architecture}
\label{sec:tacan_architecture}

\begin{figure}[t!]
\centering
\includegraphics[width=1\columnwidth]{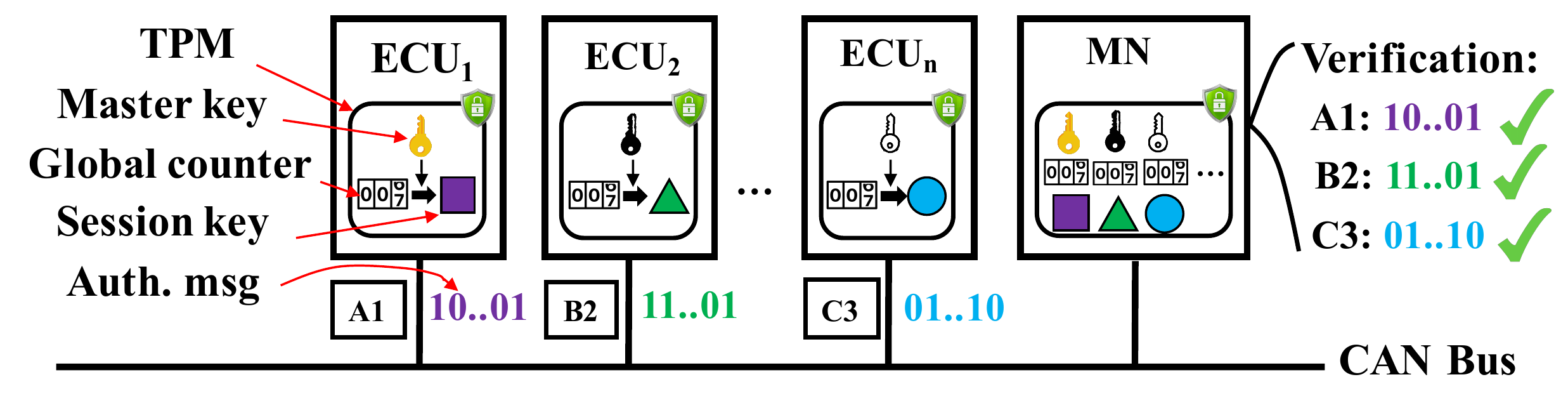}
\vspace{-0.6cm}
\caption{Illustration of TACAN architecture and the transmitter authentication protocol.}
\label{fig:TACAN_arch}
\vspace{-0.2cm}
\end{figure}

As shown in Figure~\ref{fig:TACAN_arch}, TACAN consists of in-vehicle ECUs and a trusted MN connected to the same CAN bus.
We assume that the MN is pre-installed by the manufacturer during production and requires direct physical access by authorized parties (e.g., an authorized repairs shop) to prevent potential tampering and compromises.
We further assume that the deployed covert channels are pre-configured during production or re-configured during maintenance, so that the one-way communication of authentication information from ECUs to the MN can be successful established.

Similar to~\cite{hazem2012lcap,radu2016leia,herrewege2011canauth}, a master key (MK) is assumed to be pre-shared between each ECU and the MN, which is stored in the TPM. 
Updating of MKs (e.g., when adding or replacing an ECU) should again require direct physical access by authorized parties to the involved ECUs.
The procedure of key updating is outside the scope of this paper.
During operation, Each ECU will generate a session key (SK) from the MK and a global counter and further use it to generate authentication messages.
We describe the transmitter authentication protocol in more detail in the following section.

\subsection{Transmitter Authentication Protocol}
\label{sec:transmission_authentication_protocol}
Inspired by the work in~\cite{radu2016leia}, \shaun{the MN performs unidirectional authentication of each transmitting ECU.}
\shaun{The parameters used by TACAN \rev{for the $n$ ECUs} are summarized as follows:}
\begin{itemize}
    \item The master key $MK_i$ with $i \in \{1,...,n\}$ is a pre-shared key between $ECU_i$ and the MN, which is securely stored in the TPMs of both parties. 
    
    \item The session key $SK_i$ with $i \in \{1,...,n\}$ is used to generate authentication messages for $ECU_i$.
    
    \item \rev{The local counter $l_i$ with $i \in \{1,...,n\}$ is an incremental value that stores the number of transmitted authentication messages \shaun{for $ECU_i$}. 
    This value is contained in the authentication message.} 
    
    \item The global counter $g_i$ \rev{with $i \in \{1,...,n\}$} represents a value updated on specific circumstances (e.g., car ignition, $l_i$ overflow) and it is used to generate $SK_i$. 
    
\end{itemize}
We assume that \shaun{both $SK_i$ and $g_i$} are stored in the TPM so that an attacker cannot tamper with them and launch replay attacks.

\vspace{0.1cm}
\noindent \textbf{Session Key Generation.}
Each $ECU_i$ stores its own master key $MK_i$, and the MN stores the master keys of all ECUs.
The session key for 
\rev{the $n$ ECUs} 
is generated from $MK_i$ and $g_i$ as follows:
\begin{equation}
    SK_i = HMAC(MK_i,g_i), \text{ for } i \in \{1,...,n\}, \nonumber
\end{equation}
\rev{where $HMAC(\cdot)$ refers to the Hash-based Message Authentication Code algorithm \cite{krawczyk1997hmac}.
It is possible to use different hashing algorithms for HMAC (e.g., HMAC-SHA256) to meet the desired security requirements. 
}

Every time $SK_i$ is updated, $l_i$ will be reset to zero.
On the receiver side, the MN uses the same $MK_i$ and $g_i$ to compute the corresponding $SK_i$ for $ECU_i$.
Counter synchronization is performed when $g_i$ is incremented. 
Since authentication message verification failures can be caused by de-synchronization of counters, a re-synchronization procedure may be performed. 
More details are provided in~\cite{radu2016leia}.

\vspace{0.1cm}
\noindent \textbf{Authentication Message Generation.}
$ECU_i$ first increments $l_i$ and then computes the authentication message $A_m$ as follows: 
\begin{equation}
    A_{m} = l_i || HMAC(SK_i, l_i), \nonumber 
\end{equation}
where \rev{``$||$'' denotes bit string concatenation}. 
For the scope of our work, we assume that all the parameters (keys and counters) conceived for TACAN are binary values.
As we can see, $A_{m}$ is not related to any normal CAN message and merely serves as an identifier for $ECU_i$. 

As for $l_i$, a 24-bit counter can last for $46+$ hours for a 10-ms message even in per-message authentication, which is sufficient for our transmitter authentication protocol that transmits $A_{m}$ at a much smaller frequency. 
\rev{Implementations of TACAN are free to use whichever hashing algorithm for HMAC and sizes of keys that are deemed strong enough.}
Instead of transmitting the entire digest that are usually hundreds of bits long, the transmitter may truncate each digest to several bits to reduce the transmission time (e.g., using the least significant 8 bits or XORing all bytes together to create a condensed 8-bit version of the digest, as in~\cite{kurachi2014cacan,szilagy2008flexible}).

\subsection{IAT-Based Covert Channel}
\label{sec:IAT_based_covert_channel}
\begin{figure}[t]
    \centering
    \includegraphics[trim=0cm 0.3cm 0cm 0cm,clip=true,width=1\columnwidth]{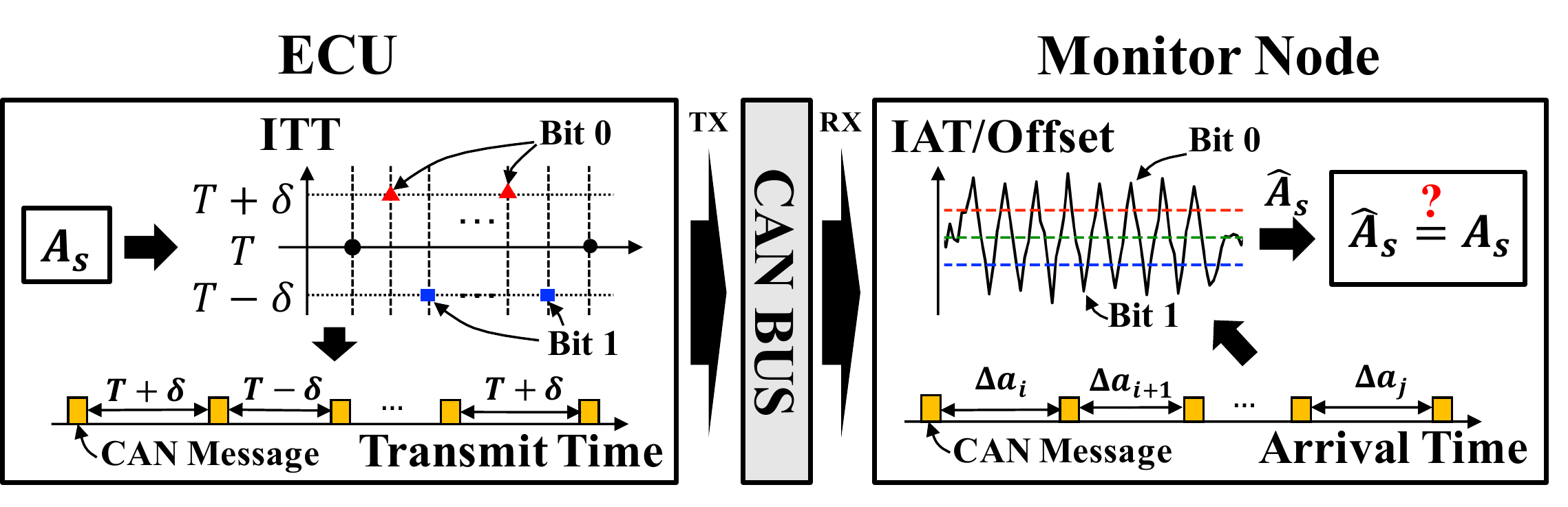}
    \caption{Illustration of timing-based covert channels.}
    \label{fig:timing_based_scheme}
    \vspace{-0.3cm}
\end{figure}

Figure~\ref{fig:timing_based_scheme} illustrates two timing-based covert channels \rev{for periodic CAN messages}, in which the transmitting ECU embeds the authentication message into the ITTs of CAN messages, which can be extracted from the IATs or offsets by the MN. 
By verifying the received authentication message, the MN can authenticate the transmitter.
\shaun{In this section, we present the IAT-based covert channel.}

\vspace{0.1cm}
\noindent \textbf{Observations.}
According to our timing model in Eq.~(\ref{eq:arrival_timestamp}), the observed IAT between messages $(i-1)$ and $i$ is given by 
\begin{equation}
    \Delta a_i = a_i - a_{i-1} = \frac{1}{1+S} \Delta t_i + (\eta_i - \eta_{i-1}). \label{eq:inter_arrival_time}
\end{equation}
Since almost all CAN messages are periodic (i.e., $\Delta t_i = T$), the IATs have a mean of $\mathbb{E}[\Delta a_i] = T/(1+S)$ and a variance of $\text{Var}(\Delta a_i) = 2\sigma_\eta^2$. 
In practice, clock skews are usually very small (in the order of 100s of ppm).
Hence, we have $\kappa = \mathbb{E}[\Delta a_i] \approx T$, where $\kappa$ is considered as the reference level for IATs\footnote{In the case of very large clock skews, the reference level $\kappa$ needs to be calibrated offline from the training dataset. }.  

From Eq.~(\ref{eq:inter_arrival_time}), we can see that an amount of deviation $\delta$ in ITTs will lead to a corresponding change of $\delta/(1+S)$ in IATs, which may be easily observed by the receiver when variances in IATs are small. 
In the case of large variances in IATs, the receiver may compute the running average to smooth out the noise, that is,
\begin{equation}
    \Delta \bar{a}[i] = \frac{1}{L} \sum_{j=0}^{L-1}\Delta a_{i+j} = \frac{1}{1+S} \left( \frac{1}{L}  \sum_{j=0}^{L-1}\Delta t_{i+j} \right) + \frac{1}{L}(\eta_{i+L-1} - \eta_{i-1}), \nonumber
\end{equation}
where $L$ is the window length. 
Through the running average, the variance of IATs can be reduced significantly by a factor of $L^2$. 
Note that due to the existence of $1/L$ in the bracket, the amount of deviation $\delta$ needs to be added to $L$ consecutive ITTs in order to maintain the same level of changes in observed IATs. 

\begin{figure}[t!]
    \centering
    \begin{subfigure}[b]{0.48\columnwidth}
        \includegraphics[trim=0cm 0.3cm 0cm 0cm,clip=true,width=\textwidth]{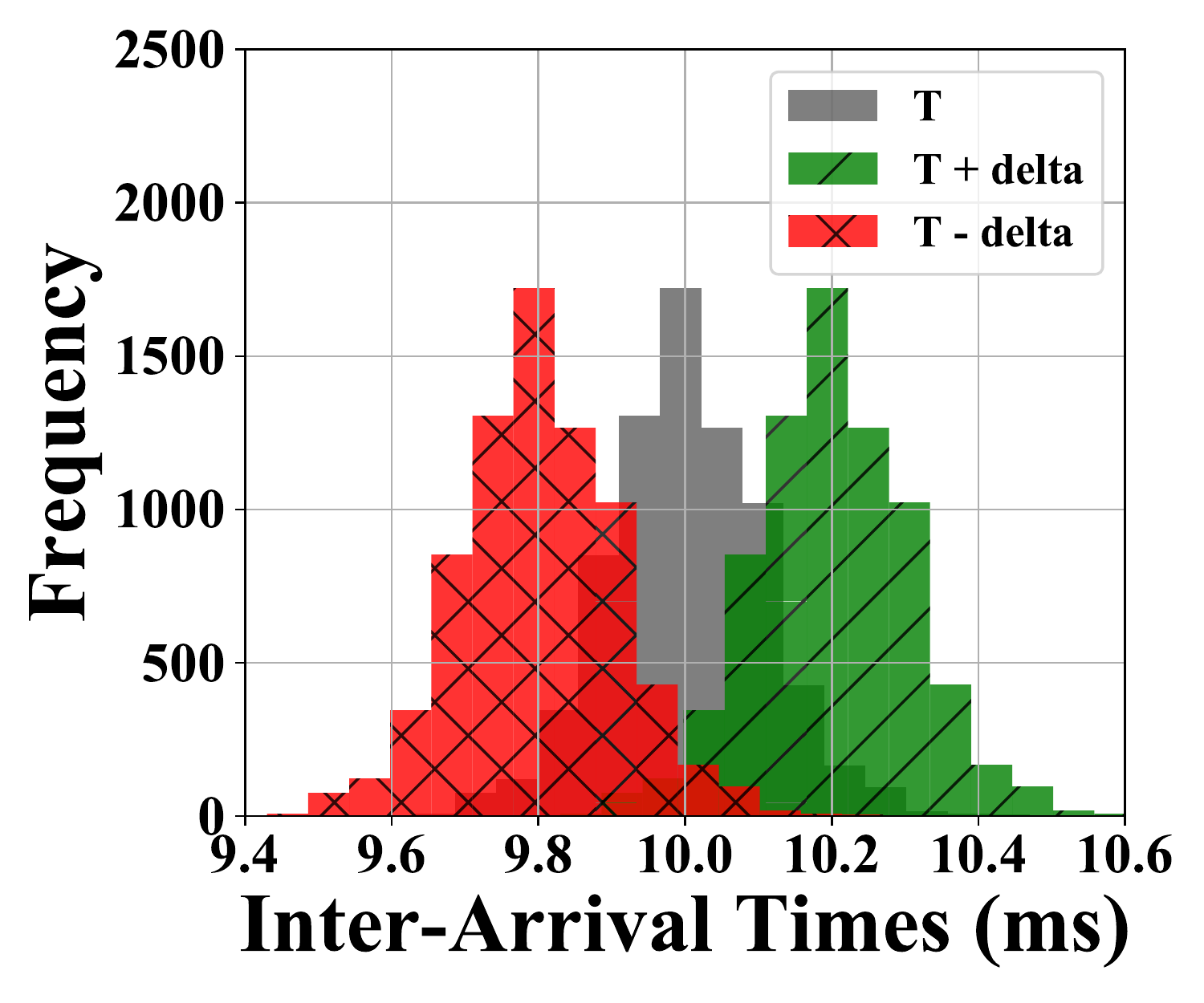}
        \caption{}
        \label{fig:example_iat_distribution_L-1}
    \end{subfigure}
    ~
    \begin{subfigure}[b]{0.48\columnwidth}
        \includegraphics[trim=0cm 0.3cm 0cm 0cm,clip=true,width=\textwidth]{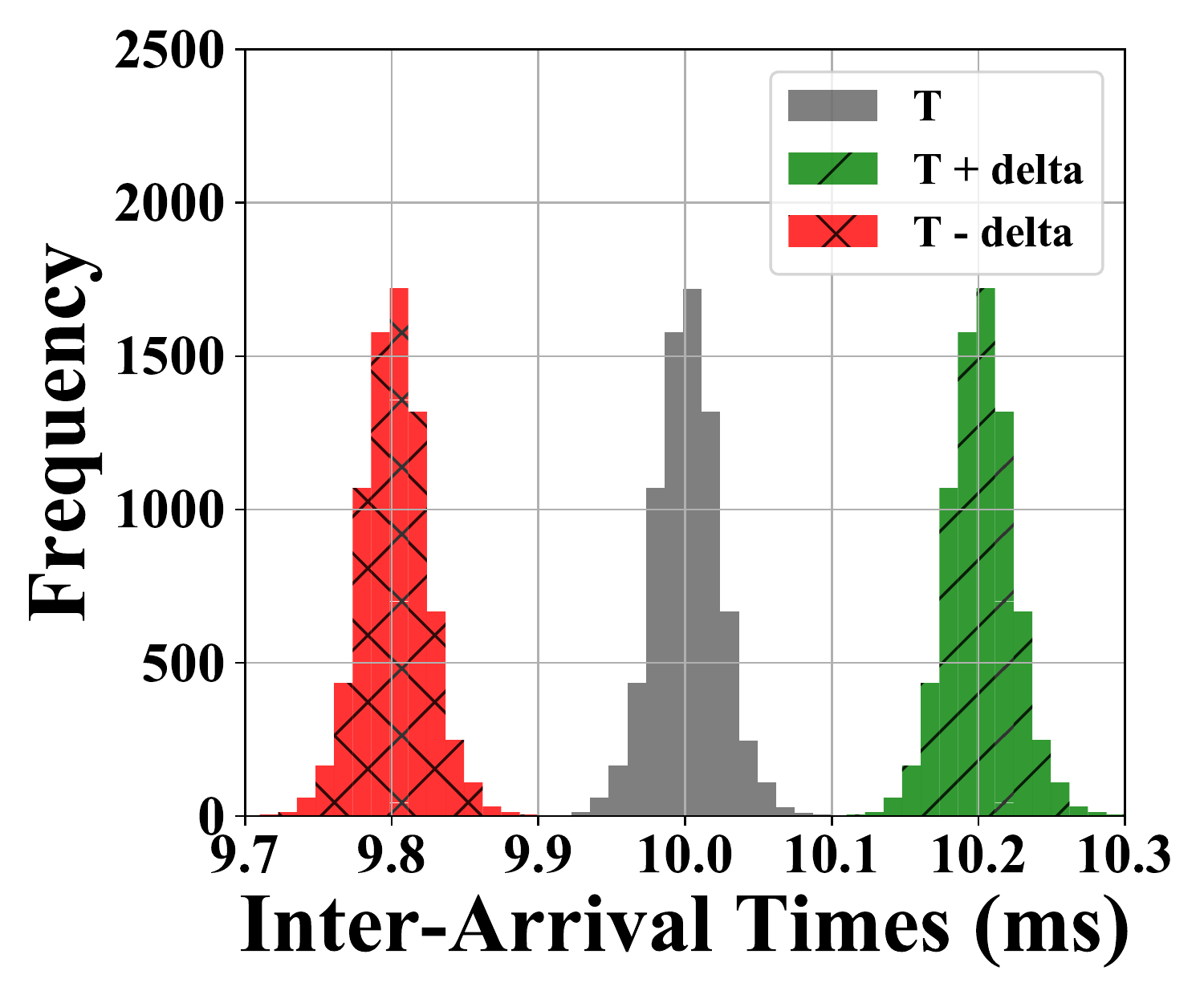}
        \caption{}
        \label{fig:example_iat_distribution_L-4}
    \end{subfigure}
    \vspace{-0.2cm}
    \caption{Example of IAT distributions of message ID=0x020 from the Toyota dataset (a) without running average ($L=1$) and (b) with running average ($L=4$).} 
    \label{fig:example_iat_distribution}
     \vspace{-0.2cm}
\end{figure}

As an example, we plot the IAT distributions of message 0x020 ($T=0.01$ sec) from the Toyota dataset~\cite{toyota2010dataset} in Figure~\ref{fig:example_iat_distribution}. 
When $\delta=0.02T=2\cdot 10^{-4}$ sec is added to or subtracted from IATs (to simulate operations at the transmitter assuming negligible clock skew effects), the three clusters (representing a bit 0 or a bit 1 or neither) cannot be separated from each other (Figure~\ref{fig:example_iat_distribution_L-1}), which may lead to bit errors at the receiver side. 
In contrast, with the running average of $L=4$, the clusters are clearly distinguishable (Figure~\ref{fig:example_iat_distribution_L-4}) and can be separated through thresholding.
The above observations motivate our design of IAT-based covert channels.


\vspace{0.1cm}
\noindent \textbf{Embedding $A_m$ into ITTs.}
Given $A_m$ of $n_m$ bits, the transmitter first constructs an authentication frame $A_f$ by inserting $n_s/2$ silence bits before and after $A_m$. 
The main purpose of silence bits (setting $\delta=0$) is to maintain the reference level of IATs and signal the start and end of $A_m$.
Hence, the total length of $A_f$ is $n_f=n_m+n_s$. 
For instance, if $A_m=$ {\tt 0101} and $n_s = 2$, we have $A_f=$ {\tt \_0101\_}.  
Each bit $i$ is modulated into $L$ consecutive ITTs as follows,
\begin{equation*}
    \Delta t_j =
    \begin{cases}
        T + \delta, & \text{ if } b_i = 0,\\
        T - \delta, & \text{ if } b_i = 1,\\
        T, & \text{ else},
    \end{cases}
\end{equation*}
\rev{where $j \in [iL, (i+1)L)$.}

\vspace{0.1cm}
\noindent \textbf{Extracting $A_m$ from IATs.}
On the receiver side, the MN records the arrival timestamps for the targeted messages and computes the \rev{IATs}. 
It then performs running average with the right choice of $L$ to obtain $\{\Delta \bar{a}[i]\}$. 
Since each bit is repeated for $L$ consecutive ITTs, the receiver needs to sample $\{\Delta \bar{a}[i]\}$ every $L$ values.
Let the sampling offset be $\tau$ and the $j$-th sample be $\Delta \bar{a}[jL+\tau]$. 
If $\tau$ is correctly chosen, the total distance between each modulated sample and the reference level ($\kappa$) should be maximized, i.e.,
\begin{equation}
    \tau^* = \arg \max_\tau \sum_j \left|\Delta \bar{a}[jL+\tau] - \kappa \right|. \nonumber
\end{equation}
Then the receiver can convert the sampled values to bits through thresholding as follows, 
\begin{equation}
    \hat{b}_j =
    \begin{cases}
        0, & \text{ if } \Delta \bar{a}[jL+\tau^*] > \Gamma_u,\\
        1,  & \text{ if } \Delta \bar{a}[jL+\tau^*] < \Gamma_l,\\
        \text{{\tt \_}}, 
        & \text{ else },
    \end{cases}
\end{equation}
where $\Gamma_l = \kappa - \delta/2$ and $\Gamma_u = \kappa + \delta/2$ are the lower and upper thresholds, respectively, \rev{$\kappa \approx T$, and ``\_'' is the silence bit}. After that, the output bits are concatenated and then split at the silence bits to obtain $A_m$.


Note that in order to establish a reliable IAT-based covert channel, the parameters including $L$ and $\delta$ need to be pre-shared between the transmitting ECU and the MN, as assumed in Section~\ref{sec:system_model}.
We leave online estimation of the covert channel parameters from the MN side as future work.


\subsection{Offset-Based Covert Channel}
\label{sec:offset_based}
The idea of the offset-based covert channel is very similar to the IAT-based covert channel (Figure~\ref{fig:timing_based_scheme}).
The main difference is that the former performs running average on IATs (including the added deviations) to smooth out the noise, while the later aims to accumulate the deviations to differentiate the modulated samples from noise. 
Our observations about clock offsets are as follows. 

\vspace{0.1cm}
\noindent \textbf{Observations.}
According to our timing model in Section~\ref{sec:system_model} and Eq.~(\ref{eq:arrival_timestamp}), we have $t_i=\sum_{j=1}^i \Delta t_j$, $a_0=\eta_0$, and $a_i=t_i/(1+S)+\eta_i$. 
Since the receiver only knows the nominal period ($T$) of the targeted message, it computes the observed clock offset as the difference between the expected elapsed time (up to message $i$) at the transmitter and the actual elapsed time at the receiver, i.e.,
\begin{equation}
    \hat{O}_i = iT - (a_i - a_0) = iT - \frac{1}{1+S} \sum_{j=1}^i \Delta t_j + \eta_0 - \eta_i. \nonumber
\end{equation}
As we can see, if the transmitter adds $\delta$ to $L$ consecutive ITTs ($\Delta t_j = T+\delta$ for $j=1,...,L$), the deviations will accumulate and lead to a decrease of $\delta L/(1+S)$ in $\hat{O}_i$.
Hence, by monitoring the changes in $\hat{O}_i$, the receiver can determine the transmitted bits and extract $A_m$ from clock offsets. 
The above observations motivate our design of offset-based covert channels. 



\vspace{0.1cm}
\noindent \textbf{Embedding $A_m$ into ITTs.}
Different from the IAT-based covert channel (Section~\ref{sec:IAT_based_covert_channel}), the transmitter embeds each bit of $A_f$ into $L$ consecutive ITTs as follows,
\begin{equation*}
    \Delta t_j =
    \begin{cases}
        T - \delta, & \text{ if } b_i = 0,\\
        T + \delta, & \text{ if } b_i = 1,\\
        T, & \text{ else},
    \end{cases}
\end{equation*}
\rev{for $j \in [iL, (i+1)L),$} and 
\begin{equation*}
    \Delta t_j =
    \begin{cases}
        T + \delta, & \text{ if } b_i = 0,\\
        T - \delta, & \text{ if } b_i = 1,\\
        T, & \text{ else},
    \end{cases}
\end{equation*}
\rev{for $j \in [iL+L/2, (i+1)L)$}, where $L$ is assumed to be an even integer. 
In other words, in order to transmit a bit 0/1, the transmitter adds $-\delta$/$\delta$ to the first $L/2$ ITTs and then subtracts $-\delta$/$\delta$ from the following $L/2$ ITTs so that the observed clock offset returns to the reference level after the transmission of each bit.

\vspace{0.1cm}
\noindent \textbf{Extracting $A_m$ from Offsets.}
On the receiver side, the monitor node records the arrival timestamps and processes the IATs in batches. 
Since each $A_f$ has $n_f$ bits and each bit is modulated into $L$ consecutive ITTs, a total number of $N = n_f L$ consecutive IATs belong to the same $A_f$, where $N$ is referred to as the batch size. 

Denote the $i$-th IAT in the $k$-batch as $\Delta a_{k,i}$, where $1 \leq i \leq N$.  
Then the observed clock offset with respect to the beginning of the current batch up to the $i$-th IAT is 
\begin{equation*}
    \hat{O}_{k}[i] = iT - \sum_{j=1}^{i} \Delta a_{k,j}.
\end{equation*}
Let $\kappa = (\max(\hat{O}_{k}[i]) + \min(\hat{O}_{k}[i]))/2$ be the midpoint (assuming that at least a bit 0 and a bit 1 are transmitted), which is considered as the reference level of clock offsets.
\rev{Since clock skew is usually very small (100s of ppm) and we are computing batch-wise clock offsets, the impact of clock skew is small, and thus we assume that the $\kappa$ is constant for the duration of a batch\footnote{In the case of very large clock skew (e.g., 1000s of ppm), there will be a linear trend due to clock skew in batch-wise clock offsets and thus the reference level may not be constant. We leave the detrending process of clock offsets as future work.}.}
Since each bit affects $L$ IATs, the receiver needs to sample $\{\hat{O}_{k}[i]\}$ every $L$ values with a sampling offset of $\tau$ and obtain the $j$-th sample as $\hat{O}_{k}[jL + \tau]$. 
If $\tau$ is correctly chosen, then the total distance between each sample and the reference level should be maximized, i.e.,
\begin{equation*}
    \tau^* = \arg \max_{\tau} \sum_j | \hat{O}_{k}[jL + \tau] - \kappa|. 
\end{equation*}
Then the receiver converts the sampled values to bits through the following threhsolding-based scheme,
\begin{equation*}
    \hat{b}_j =
    \begin{cases}
        0, & \text{ if } \hat{O}_{k}[jL + \tau^*] > \Gamma_u,\\
        1,  & \text{ if } \hat{O}_{k}[jL + \tau^*] < \Gamma_l,\\
        \text{{\tt \_}}, & \text{ else, }
    \end{cases}
\end{equation*}
where $\Gamma_l=\kappa - \frac{1}{4}\delta L$ and $\Gamma_u=\kappa + \frac{1}{4}\delta L$ are the lower and upper thresholds, respectively\rev{, and ``\_'' is the silence bit}. 
The term of $\frac{1}{4}\delta L$ is due to the fact that $\delta$ is added to (or subtracted from) $L/2$ consecutive ITTs, and thus the maximum total deviation is $\pm \frac{1}{2} \delta L$. 
The midpoints between $\kappa$ and $\kappa \pm \frac{1}{2} \delta L$ are chosen as thresholds. 
Eventually, $A_m$ is extracted by concatenating all decoded bits and splitting them at the silence bits. 

\vspace{0.1cm}
\noindent \rev{\textbf{Impact on CAN bus schedulability.} As we have shown, in both the IAT-based and offset-based covert channels, a certain amount of deviation is added to the ITTs.
As a result, it may increase the worst-case response time of CAN messages, which is defined as the longest time from the initiating event (that puts the message in the transmission queue) occurring to the message being received by the nodes that require it.
If we apply the schedulability analysis in \cite{davis2007controller} to TACAN, we can show that the effect of TACAN is equivalent to increasing the blocking delay by a constant amount of time (hundreds of $\mu$s) and increasing the message transmission time by a small percentage. 
Therefore, \rev{to achieve effective use of covert channels, TACAN parameters need to be experimentally obtained and fine tuned prior to deployment to ensure the schedulability of the CAN bus.}
A more detailed discussion is provided in Appendix~\ref{appendix:schedulability}.
}

\subsection{LSB-Based Covert Channel}
\label{sec:lsb_based}
\begin{figure}[t]
    \centering
    \includegraphics[width=1\columnwidth]{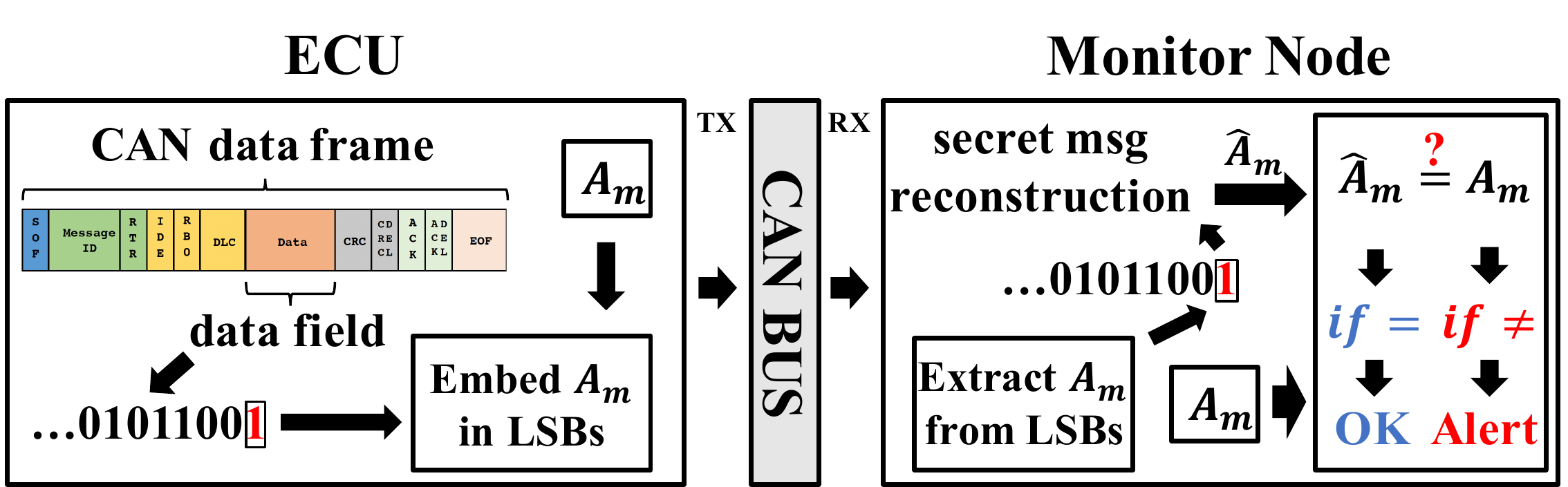}
    \caption{Illustration of LSB-based covert channel. The transmitting ECU embeds the authentication message into the LSBs of multiple normal CAN messages (with the same ID), which can be extracted and verified by the Monitor Node.}
    \label{fig:lsb_scheme}
\end{figure}

In this section, we introduce a storage-based covert channel that embeds the authentication messages inside the LSBs of the data payload of normal CAN messages transmitted by an ECU, referred to as the LSB-based covert channel (Figure~\ref{fig:lsb_scheme}).
\rev{Unlike the timing-based covert channels, the LSB-based covert channel can be applied to aperiodic CAN messages.}
For the scope of this work, we use the CAN data frames to develop our methodology.



\vspace{0.1cm}
\noindent \textbf{Observations.}
In order to transmit an authentication message over the CAN bus, it is very common to make use of the existing fields of a CAN message, such as the data field~\cite{kurachi2014cacan,hazem2012lcap,groza2012libra} (at least one byte) and the extended ID field~\cite{hazem2012lcap}, or simply introduce additional CAN messages~\cite{hazem2012lcap,groza2012libra}.
In practice, however, if all bytes in the data field have been used or the CAN bus is already heavily loaded, then the existing approaches of exploiting CAN messages may disrupt the ECU's functionality or increase arbitration delays.

Different from existing schemes that attempt to authenticate messages, our transmitter authentication scheme aims to authenticate the transmitter instead of each message.
Hence, authentication messages are transmitted much less frequently, which means that authentication bits may be spread over multiple CAN messages, using only a few bits from each CAN message.
Moreover, having realized that \rev{certain} 
CAN messages are used to carry sensor values and most of them are floating numbers, we may use the $L$ LSBs (e.g., $L=1$ or $2$) for authentication purposes without causing significant degradation in accuracy.
\rev{Taking the Toyota Camry as an example, there are at least $7$ messages out of $42$ that carry sensor values (e.g., wheel speeds, engine speed, vehicle speed, odometer, brake pressure, steering angle) \cite{toyota2010dataset}. 
We would expect more CAN messages that carry sensor values in newer automobiles.}
The above observations motivate our design of LSB-based covert channels.

\vspace{0.1cm}
\noindent \textbf{Embedding $A_m$ to LSBs.} 
\rev{The embedding process is implemented as a sub-layer between the Application and the Data Link layers. 
For each message $A_m$ (plus a known preamble to indicate the start and end of $A_m$), it substitutes the least significant $L$ bits of the CAN message with the next $L$ bits in $A_m$. 
No modification is needed if the $L$ bits happen to be the same.}

\vspace{0.1cm}
\noindent \textbf{Extracting $A_m$ from LSBs.} 
\rev{On the receiver side, the MN extracts the $L$ LSBs from the received CAN messages and reconstructs the authentication message.
If the MN fails to verify the authentication message, it will raise an alert that indicates possible compromise of the transmitting ECU or malicious exploitation of the CAN bus.}

As in the case of timing-based covert channels, we assume that the settings for the LSB-based covert channel are pre-shared between each ECU and the MN during production and updated during maintenance if necessary.
\rev{More details about the embedding and extraction processes are provided in Appendix \ref{appendix:lsb_algorithm}.}

\section{Security Discussion}
\label{sec:security_analysis}
\rev{Compared to the existing message authentication schemes \cite{herrewege2011canauth,hazem2012lcap,kurachi2014cacan,radu2016leia} that attempt to verify the authenticity of each message, TACAN extracts and verifies authentication message embedded in the timing or LSBs of normal messages to authenticate the transmitting ECU and also serves the purpose of intrusion detection.
Hence, TACAN can detect attacks that interrupt the transmission of normal CAN messages (e.g., suspension, injection, and DoS attacks), as well as attacks in which attackers fail to generate valid authentication messages (e.g., forgery, replay, and masquerade attacks). 
\rev{The above security properties provided by TACAN are independent of the CAN protocol and the contents of CAN messages.}
}

\vspace{0.1cm}
\noindent \textbf{Security Features.}
We summarize a list of features provided by TACAN as follows: 1) TACAN securely stores both master and session keys in the ECU's TPM to prevent being compromised by the adversary, 2) TACAN employs monotonic counters for generating session keys and authentication messages for each ECU, and 3) TACAN requires each ECU to continuously transmit \rev{unique} authentication messages to enable real-time transmitter authentication.

Note that due to the third feature, if the attacker compromises an in-vehicle ECU but does not attempt to deceive TACAN (i.e., stop generating authentication messages), it will interrupt the continuous transmission of authentication messages and thus will be detected immediately.
In addition, since basic attacks like \rev{suspension, injection, and DoS attacks} can be easily detected, we focus on more sophisticated attacks in the rest of this section that actively attempt to evade TACAN, including the forgery attack, the replay attack, and the masquerade attack.

\vspace{0.1cm}
\noindent \textbf{Detecting Forgery Attacks.}
In the forgery attack, the adversary has already compromised an in-vehicle ECU that is protected by TACAN and attempts to generate valid authentication messages that can be verified by the MN in order to evade the detection of TACAN. 
Since our adversary model (Section~\ref{sec:adversary_model}) assumes that the attacker has no access to the TPM of the compromised ECU, the attacker has to forge a valid digest for each local counter value without the session key.
With a condensed digest of $M$ bits, the probability of a successful forgery is $1/2^M$.
For example, when $M=8$, this probability is $1/2^8 \approx 0.4\%$.
Repeated forgeries will be prevented due to the use of monotonic counters. 

\vspace{0.1cm}
\noindent \textbf{Detecting Replay Attacks.}
A replay attacker has infiltrated the CAN bus and attempts to replay previously transmitted authentication messages of the targeted ECU with the hope of passing the verification process at the MN.
It is easy to see that such attempts will be detected by TACAN due to the use of monotonic counters.

\vspace{0.1cm}
\noindent \textbf{Detecting Masquerade Attacks.}
As mentioned in Section~\ref{sec:adversary_model}, a masquerade attack (including the more sophisticated cloaking attack \cite{sagong2018cloaking}) requires in-vehicle ECUs to be weakly and/or fully compromised.
As a result, TACAN will force the attacker to perform a forgery or replay attack, not only for the compromised ECU itself, but also for the ECU the attacker attempts to masquerade as. 
Therefore, a masquerade attack will be detected by TACAN. 

\vspace{0.1cm}


\section{Evaluation}\label{sec:evaluation}
In this section, we implement the proposed covert channels on the testbed and report their performance in terms of the throughput and bit error \rev{ratio} using the datasets collected from two real vehicles, a 2010 Toyota Camry~\cite{toyota2010dataset} and a 2016 Chevrolet Camaro (University of Washington EcoCAR)~\cite{ecocar}. Testbed validation is in Section~\ref{sec:testbed}, then the evaluations of covert channels IAT-based, offset-based, and LSB-based are in Section~\ref{sec:evaluation_of_IAT_based_covert_channel}, Section~\ref{sec:evaluation_of_offset_based}, and Section~\ref{sec:evaluation_of_lsb}, respectively.

\subsection{Testbed Validation}
\label{sec:testbed}
\begin{figure}[t!]
\centering
\includegraphics[width=1\columnwidth]{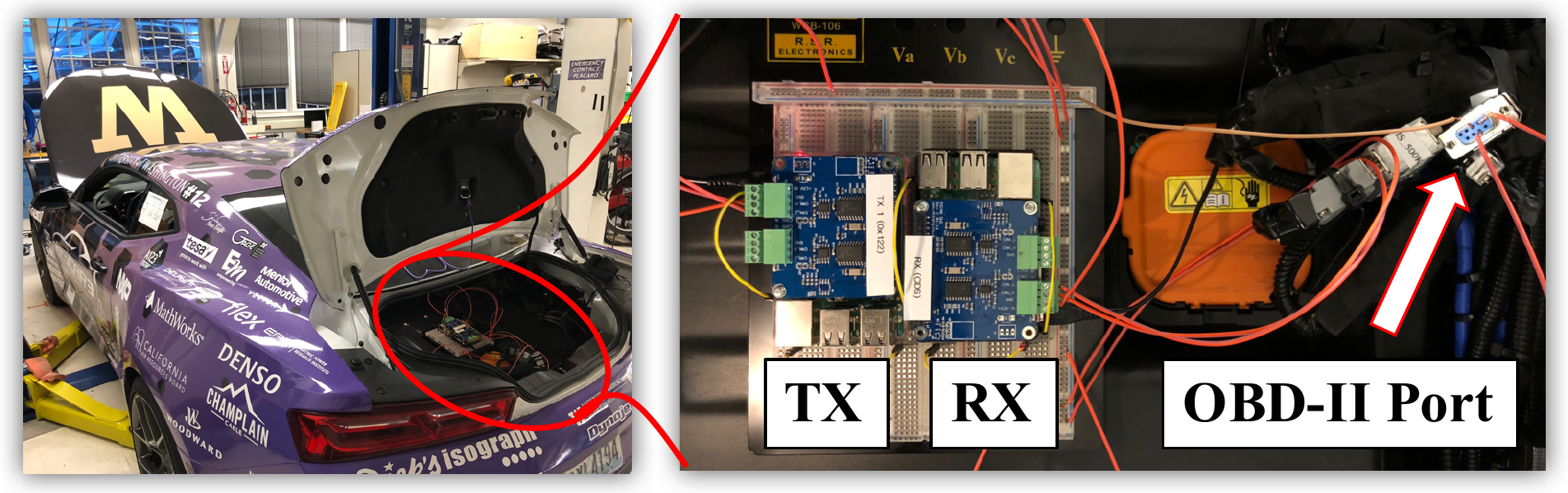}
\caption{EcoCAR testbed. Two Raspberry Pi-based ECUs are connected to the OBD-II port at the back of the EcoCAR, one as the transmitter and the other one as the receiver.}
\label{fig:testbed_setup}
\end{figure}

As shown in Figure~\ref{fig:testbed_setup}, our EcoCAR testbed consists of the UW EcoCAR and two testbed ECUs, which are connected via the On-Board Diagnostics (OBD-II) port. 
The UW EcoCAR hosts 8 stock ECus and two experimental ECUs.
There are a total of 2500+ messages using 89 different messages IDs are exchanged on the CAN bus every second.
Each ECU consists of a Raspberry Pi 3 and a PiCAN 2 board (using a MCP2515 CAN controller and a MCP2551 CAN transceiver). 
The SocketCAN \cite{socketCAN} library is used to enable the interaction between the added ECUs and the UW EcoCAR.

In order to demonstrate the proposed covert channels, we implemented them in Python in the transmitter and collected the CAN data traces using the receiver. 
\rev{Since the LSB-based covert channel is expected to have a very small bit error \rev{ratio},} we focus on the timing-based (IAT-based and offset-based) covert channels.
As an example, we set $n_m=36$ bits (length of authentication message) \rev{with alternating bit values} and $n_s=4$ bits (number of silence bits). 
Hence, each authentication frame is $n_f=40$ bits long. 
In order to avoid any conflict with existing messages, we chose a non-existent message ID of 0x180 and set the message period $T =10$ ms \rev{and $\delta=0.2$ ms ($2\%$ of $T$)}. 
\rev{While a non-existent message ID was used in our experiments, we assumed the Raspberry Pi-based ECU as a stock ECU that transmits messages with ID=0x180 and that is capable of modifying the timing of message ITTs.
We would like to emphasize that TACAN does not require a new message ID or additional CAN messages to implement the proposed covert channels.}

\begin{figure}[t!]
    \centering
    \begin{subfigure}[b]{0.48\columnwidth}
        \includegraphics[width=\textwidth]{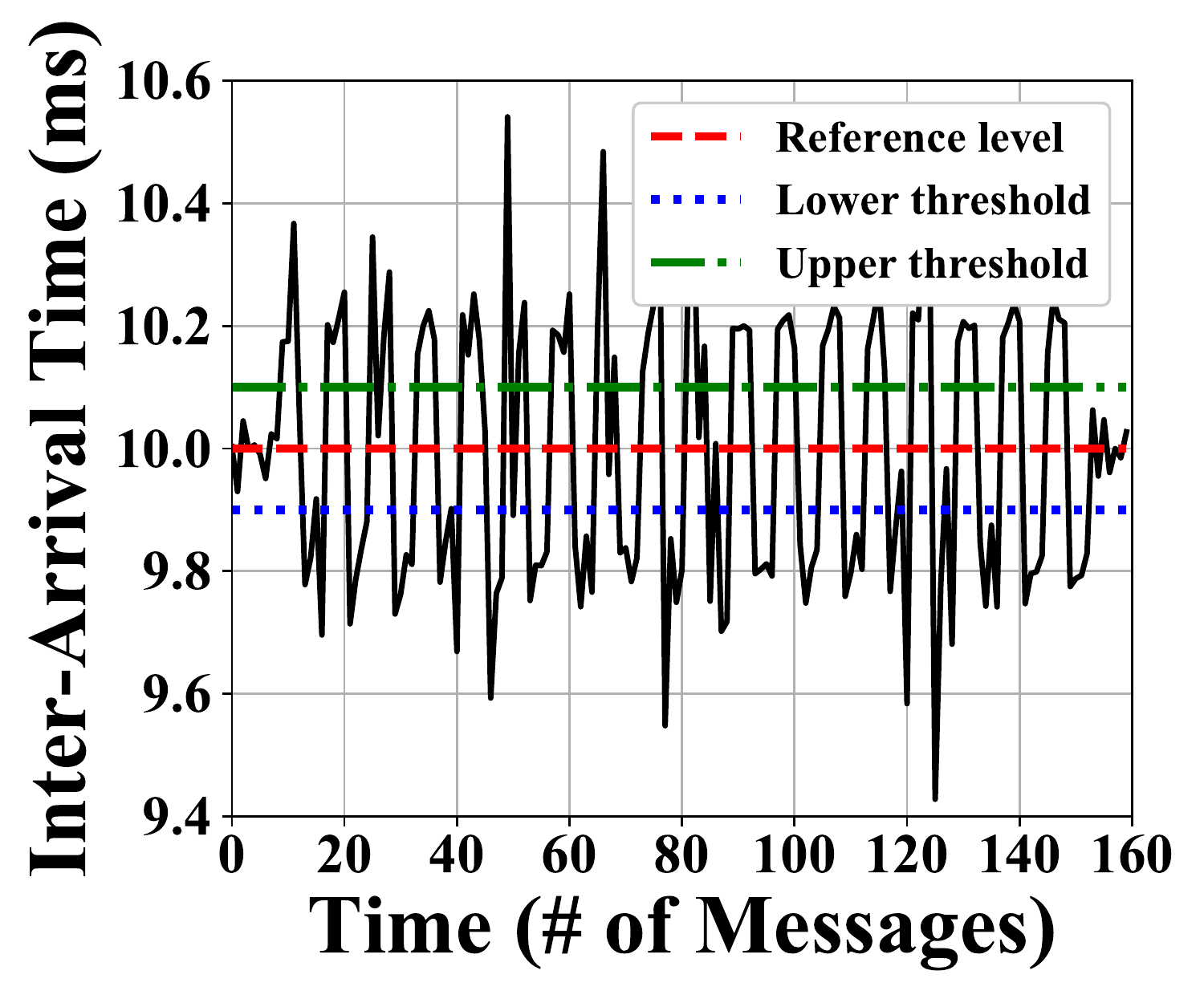}
        \caption{}
        \label{fig:testbed_validation_IAT_based_without_running_average}
    \end{subfigure}
    ~
    \begin{subfigure}[b]{0.48\columnwidth}
        \includegraphics[width=\textwidth]{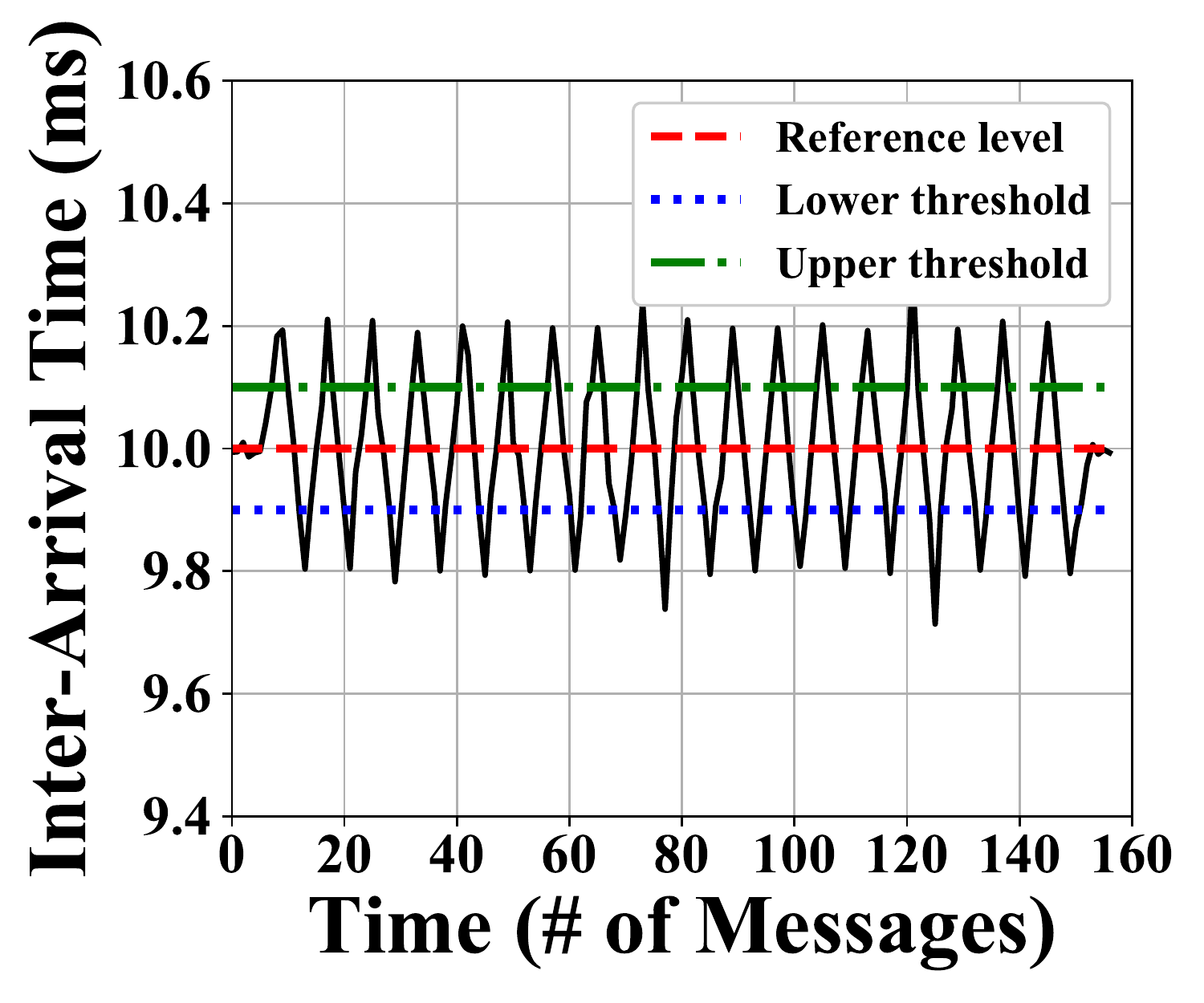}
        \caption{}
        \label{fig:testbed_validation_IAT_based_with_running_average}
    \end{subfigure}
    \vspace{-0.2cm}
    \caption{Example of observed IATs in IAT-based covert channels (a) without and (b) with running average ($L=4$).}
    \label{fig:testbed_validation_IAT_based}
\end{figure}

Figure~\ref{fig:testbed_validation_IAT_based} provides an example of observed IATs in the IAT-based covert channel. 
As we can see, without running average, the modulated IATs are noisy, and thresholding at the receiver side may lead to many possible bit errors (Figure~\ref{fig:testbed_validation_IAT_based_without_running_average}). 
In contrast, the running average process can effectively reduce IAT variations, making the peaks and valleys of the modulated IATs clearly stand out, which indicates a smaller probability of bit errors (Figure~\ref{fig:testbed_validation_IAT_based_with_running_average}).

\begin{figure}[t!]
    \centering
    \begin{subfigure}[b]{0.48\columnwidth}
        \includegraphics[width=\textwidth]{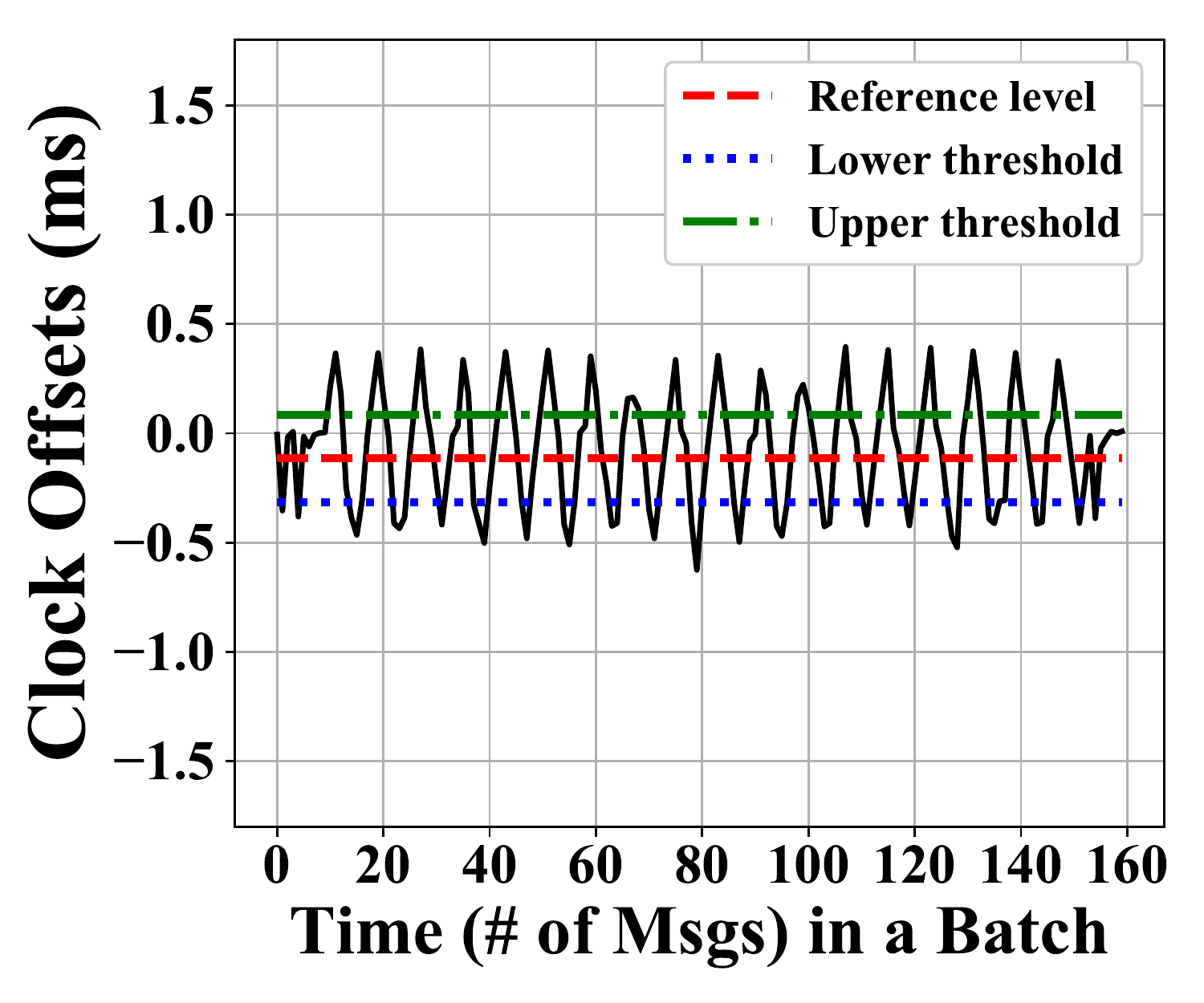}
        \caption{}
        \label{fig:testbed_validation_offset_based_L-4}
    \end{subfigure}
    ~
    \begin{subfigure}[b]{0.48\columnwidth}
        \includegraphics[width=\textwidth]{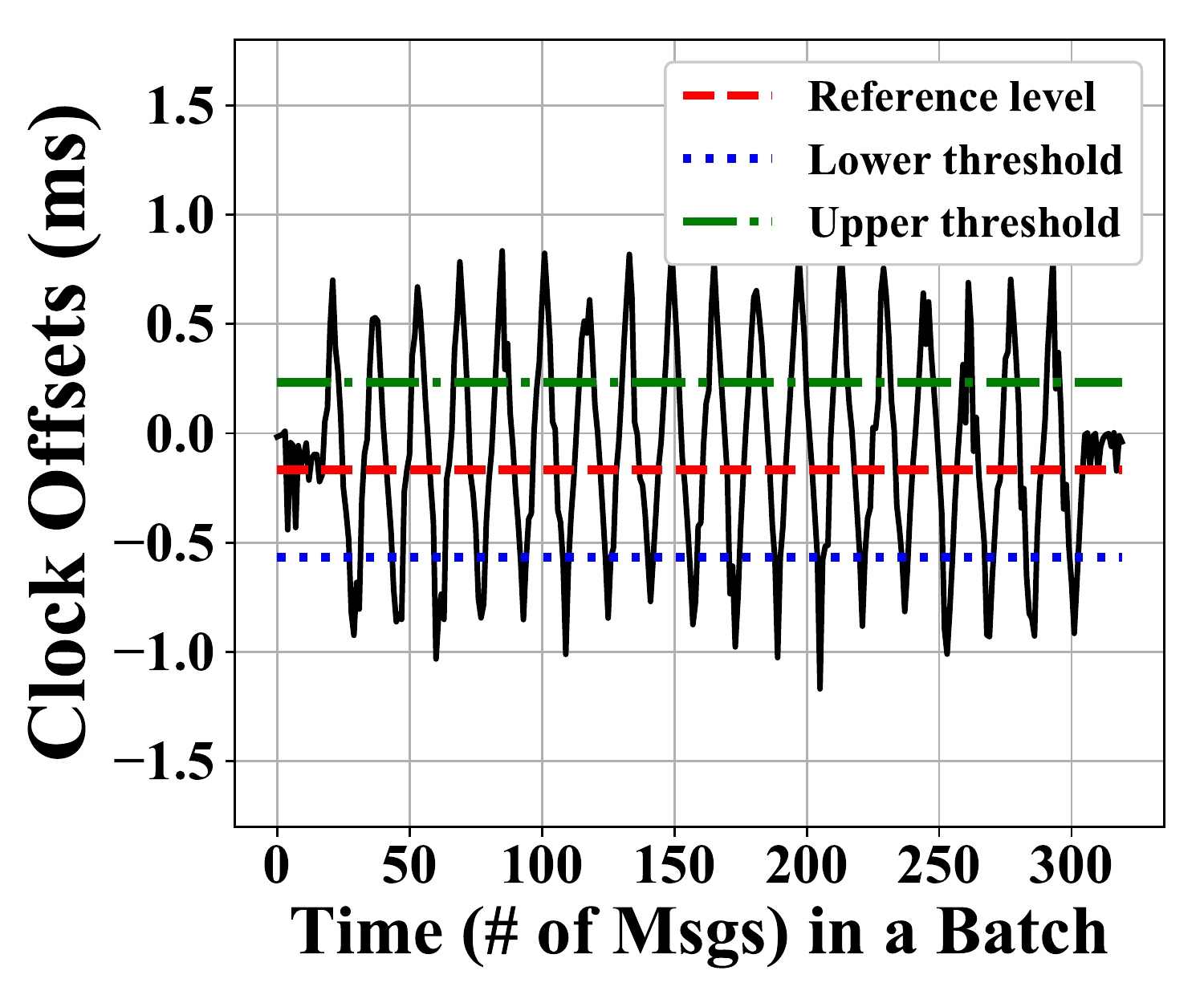}
        \caption{}
        \label{fig:testbed_validation_offset_based_L-8}
    \end{subfigure}
    \vspace{-0.2cm}
    \caption{Example of observed offsets in offset-based covert channels with (a) $L=4$ and (b) $L=8$. 
    }
    \label{fig:testbed_validation_offset_based}
    \vspace{-0.3cm}
\end{figure}

An example of observed offsets in the offset-based covert channel is provided in Figure~\ref{fig:testbed_validation_offset_based}. 
As we can see, unlike the IAT-based covert channel, increasing $L$ in the offset-based covert channel effectively increases the accumulated amount of deviations in offsets at the receiver side and thus reduces the bit error probability.
In the rest of this section, we evaluate each covert channel in more detail.

\subsection{Evaluation of IAT-Based Covert Channel}
\label{sec:evaluation_of_IAT_based_covert_channel}
In this section, we evaluate the performance of IAT-based covert channels in terms of throughput and bit error \rev{ratio}. 

\vspace{0.1cm}
\noindent \textbf{Throughput.}
Given $n_m$ bits for $A_m$, $n_s$ silence bits, and a window length of $L$, the time it takes for a CAN message with period $T$ to transmit an authentication frame $A_f$ is $T_{f} = (n_m + n_s) L T$, which increases linearly as function of $L$ and $T$. 
Then the throughput or rate for transmitting $A_m$ is $r_m = n_m/T_f$ bits per second (bps). 
For example, suppose that $n_m=36$, $n_s=4$, and $L=4$. 
Then we have $T_f = 1.6$ sec for $T=0.01$ sec and $r_m = 36/1.6 = 22.5$ bps. 

\vspace{0.1cm}
\noindent \textbf{Bit Error \rev{Ratio}. }
\rev{The bit error ratio is defined as  the number of bit errors divided by the total number of bits transmitted in a given time interval.}
In order to evaluate the bit error \rev{ratio} of IAT-based covert channels on real vehicles, we collected data for six representative messages (including message 0x180 that was transmitted from the testbed ECU) with different ID levels, periods, and noise levels from the Toyota Camry~\cite{toyota2010dataset} and the UW EcoCAR~\cite{ecocar}, as shown in Table~\ref{table:representative_messages}.
The same EcoCAR dataset was used in~\cite{ying2018shape}. 
Note that IAT noise is quantified in terms of the standard deviation (normalized by the period) and the range (the difference between the maximum IAT and the minimum IAT, normalized by the period). 

\begin{table}[h!]
    \footnotesize
	\centering
	\caption{Selected set of representative messages.}
	\label{table:representative_messages}
    \begin{tabular}{|c|c|c|c|c|}
    \hline
    \multirow{2}{*}{Msg ID} & \multirow{2}{*}{\begin{tabular}[c]{@{}c@{}}Period\\ (ms)\end{tabular}} & \multirow{2}{*}{\begin{tabular}[c]{@{}c@{}}Standard dev.\\ (Normalized)\end{tabular}} & \multirow{2}{*}{\begin{tabular}[c]{@{}c@{}}Range\\ (Normalized)\end{tabular}} & \multirow{2}{*}{Source} \\
     &  &  &  &  \\ \hline
    0x020 & 10 & 1.1\% & 10.2\% & Toyota \\ \hline
    0x224 & 30 & 0.9\% & 4.8\% & Toyota \\ \hline
    0x0D1 & 10 & 2.7\% & 51.5\% & EcoCAR \\ \hline
    0x180 & 10 & 1.7\% & 30.1\% & EcoCAR \\ \hline
    0x185 & 20 & 1.3\% & 22.6\% & EcoCAR \\ \hline
    0x22A & 100 & 1.2\% & 6.4\% & EcoCAR \\ \hline
    \end{tabular}
    \normalsize
\end{table}

In this experiment, we first preprocess the data trace of each message to fill in any missing messages.
We set $n_m=36$ bits and $n_s=4$ bits, and thus each frame is $40$ bits long.
Note that the choices of $n_m$ and $n_s$ are not critical here, since we focus on the bit error \rev{ratio} instead of the frame error \rev{ratio}. 
Out of $100$ frames ($100 n_m$ bits in total), the number of bit errors ($\hat{b}_j \neq b_j$) is recorded and divided by the number of transmitted bits to obtain the bit error \rev{ratio}. 
The amount of deviation $\delta$ is fixed to $2\%$ of the message period ($T$). 
Since clock skews are small, we directly add $\delta$ to IATs to simulate the effect of adding the same amount to ITTs at the transmitter side.

\begin{figure}[t!]
\centering
\includegraphics[width=.7\columnwidth]{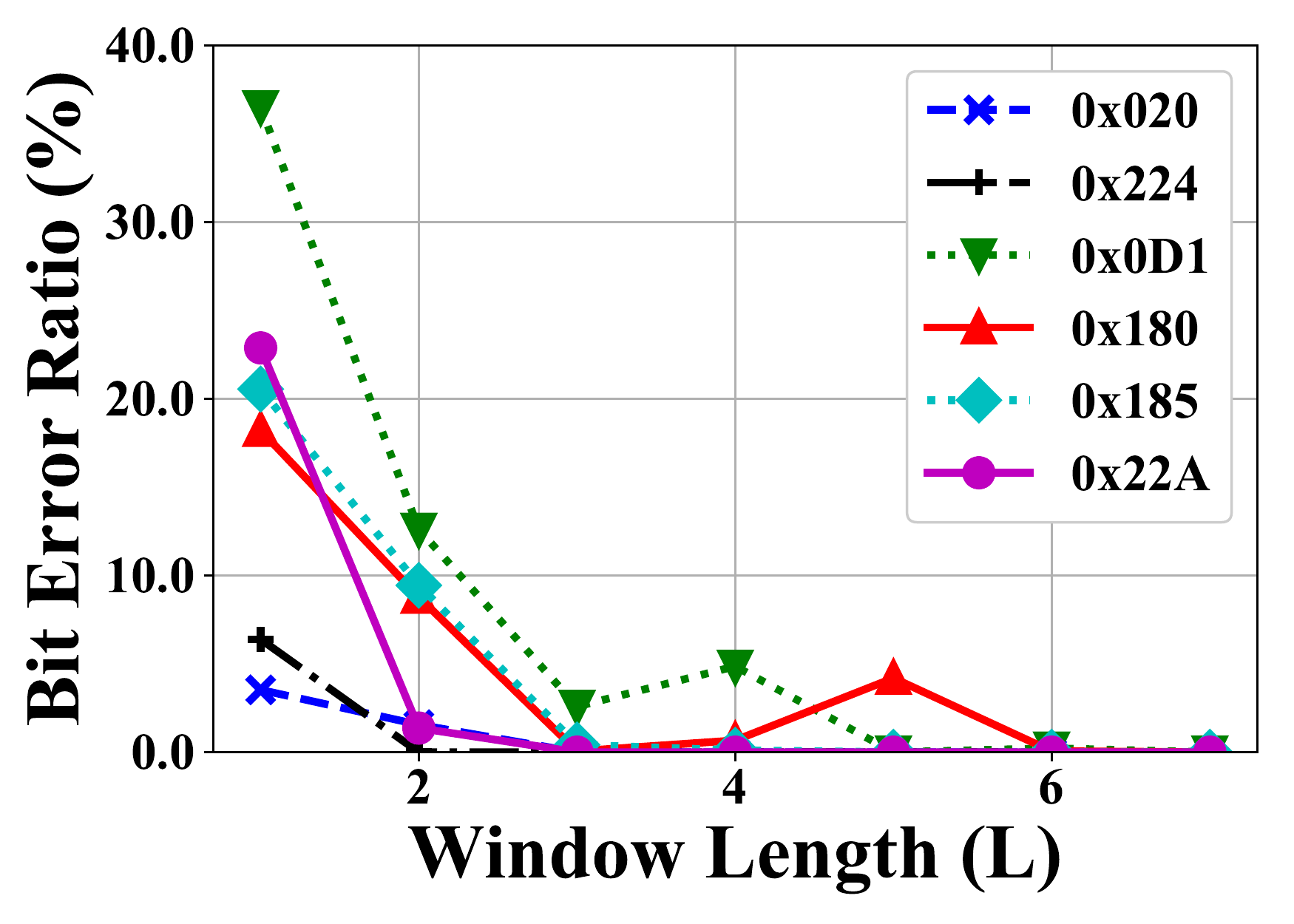}
\caption{Bit error \rev{ratios} of IAT-based covert channel for different CAN messages as a function of window length (L).}
\label{fig:evaluation_IAT_based_covert_channel}
\end{figure}

As illustrated in Figure~\ref{fig:evaluation_IAT_based_covert_channel}, the performance of IAT-based covert channel varies a lot among different messages, depending on their noise levels. 
In general, the bit error \rev{ratios} are very high with $L=1$, but they quickly drop to $0$ as $L$ increases for all messages, which demonstrates the effectiveness of running average. 
Moreover, we observe that messages with a large IAT range (e.g., 0x180, 0x0D1 and 0x185) tend to require a large $L$ value to establish a reliable IAT-based covert channel. 
With $L=6$, all messages have a bit error \rev{ratio} of less than $0.1\%$.

\subsection{Evaluation of Offset-Based Covert Channel}
\label{sec:evaluation_of_offset_based}

\begin{figure}[t!]
\centering
\includegraphics[width=.7\columnwidth]{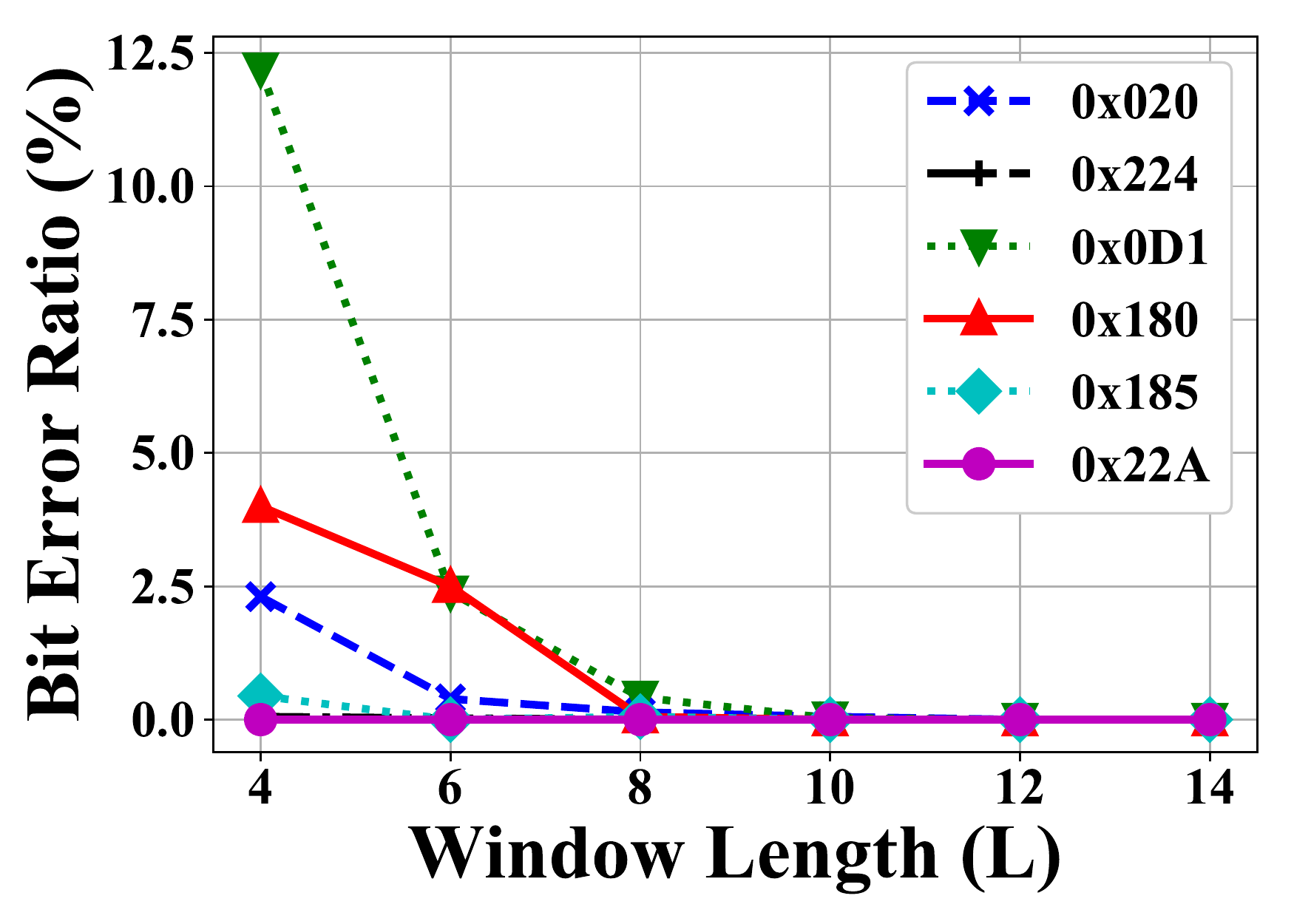}
\caption{Bit error \rev{ratios} of offset-based covert channels for different CAN messages as a function of window length (L).}
\label{fig:evaluation_offset_based_covert_channel}
\end{figure}

Since the throughput analysis of offset-based covert channels is the same with that of IAT-based covert channels, we focus on the bit error performance in this experiment using the same setup in Section~\ref{sec:evaluation_of_IAT_based_covert_channel}.
Results are provided in Figure~\ref{fig:evaluation_offset_based_covert_channel}. 

As we can see, the performance of offset-based covert channels again depends on the characteristics of individual messages. 
When $L=4$, the bit error \rev{ratio} is less than $1\%$ for messages 0x224, 0x185, and 0x22A. 
As $L$ increases, added deviations accumulate at the receiver side, thus effectively reducing the bit error \rev{ratio}. 
With $L=8$, the bit error \rev{ratio} has dropped to less than $0.42\%$ for all messages. 
As compared to the IAT-based covert channel, the offset-based covert channel is less efficient and requires a larger $L$ value.
This is mainly because it adds deviations to only the first half of $L$ consecutive ITTs, and the second half is used only for the purpose of maintaining the reference level for clock offsets, which contributes very little in differentiating the modulated samples from the noise. 

\subsection{Evaluation of LSB-Based Covert Channel}
\label{sec:evaluation_of_lsb}
In this section, we evaluate the performance of the LSB-based covert channel in terms of throughput, bit error \rev{ratio}, and accuracy loss.

\vspace{0.1cm}
\noindent \textbf{Throughput.}
Given $n_m$ bits for $A_m$ and $n_s$ bits for the start sequence, it takes $T_f=n_f T/L = (n_m+n_s)T/L$ sec to transmit an authentication frame $A_f$ using $L$ LSBs of a CAN message with period $T$.
Then the throughputs for transmitting $A_f$ and $A_m$ are $r_f=n_f/T_f$ and $r_m=n_m/T_f$, respectively.
For example, suppose that $n_s=36$, $n_s=4$, $T=0.01$ sec and $L=2$. 
Then we have $T_f=0.2$ sec, $r_f=40/0.2=200$ bps, and $r_m=36/0.2=180$ bps. 


\vspace{0.1cm}
\noindent \textbf{Bit Error \rev{Ratio}.}
Since the LSB-based covert channel embeds the authentication message into the data payload of normal CAN messages, its bit error \rev{ratio} will be as small as that of the CAN bus itself. 
According to~\cite{ferreira2004experiment}, the bit error \rev{ratios} of CAN in normal environments (factory production line) and aggressive environments (two meters away from a high-frequency arc-welding machine) are $3.1\times 10^{-7}\%$ and $2.6 \times 10^{-5}\%$, respectively.


\vspace{0.1cm}
\noindent 
\textbf{Accuracy Loss.}
As the LSB-based covert channel modifies the LSBs in the data fields of CAN data frames, \rev{such modification will lead to accuracy loss of sensor values.}
\rev{With changes of one LSB ($L=1$), the accuracy loss of the same scale with the resolution (or the discretization error) of that sensor value.
Increasing $L$ will result in larger throughput and accuracy loss.
Hence, manufacturers will need to assess the impact of accuracy loss in CAN data on the functionality and safety, as well as trade off the accuracy loss against throughput when deploying the LSB-based covert channel.}
\rev{It is important to highlight that for periodic messages that cannot tolerate accuracy loss, manufacturers may deploy IAT-based and offset-based covert channels instead.}

\rev{In order to demonstrate the feasibility of the proposed LSB-based covert channel}, we consider two CAN messages: 1) the wheel velocity value based on the publicly available Toyota dataset~\cite{toyota2010dataset} and 2) the engine coolant temperature value \rev{that we identified through reverse engineering from the EcoCAR dataset~\cite{ecocar}.}
%
%
%
In our experiments, we set $L$ to $1$ or $2$ and \rev{quantify the accuracy loss in terms of \textit{the maximum error} (using the original values as ground truth)}. 
Note that we intentionally keep $L \leq 2$ in order to avoid significant distortions to the underlying sensor values or jeopardize the functionality of the receiving ECU.

\begin{figure}[t!]
    \centering
    \includegraphics[trim=-0.5cm 0.7cm 0cm 0cm,clip=true,width=.74\columnwidth]{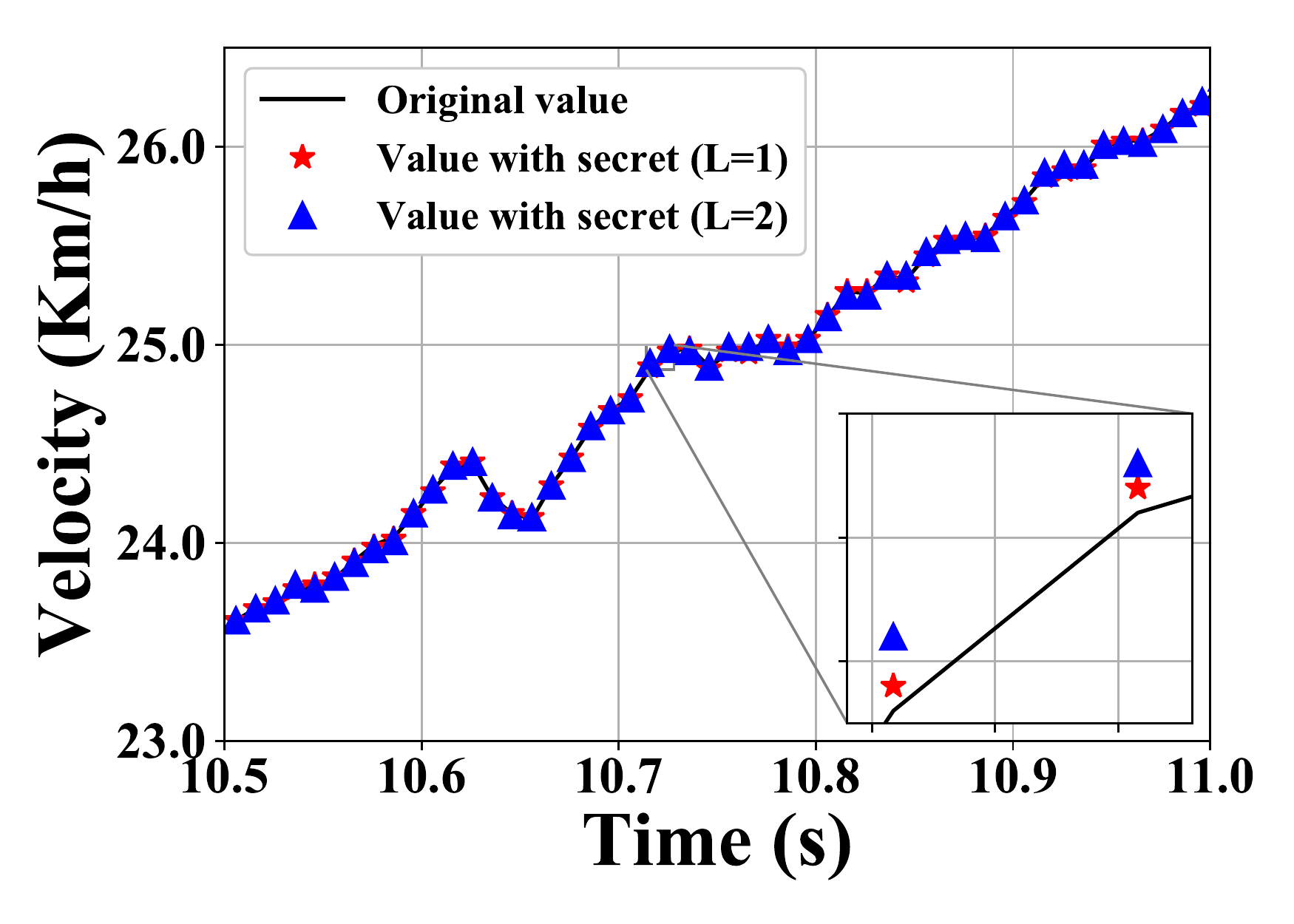}
    \caption{Authentication message embedded in Toyota wheel velocity data with $L=1$ and with $L=2$.}
    \label{fig:toyota_lsb_evaluation}
\end{figure}
\begin{figure}[t!]
    \centering
    \includegraphics[trim=0cm 0.7cm 0cm 0cm,clip=true,width=.7\columnwidth]{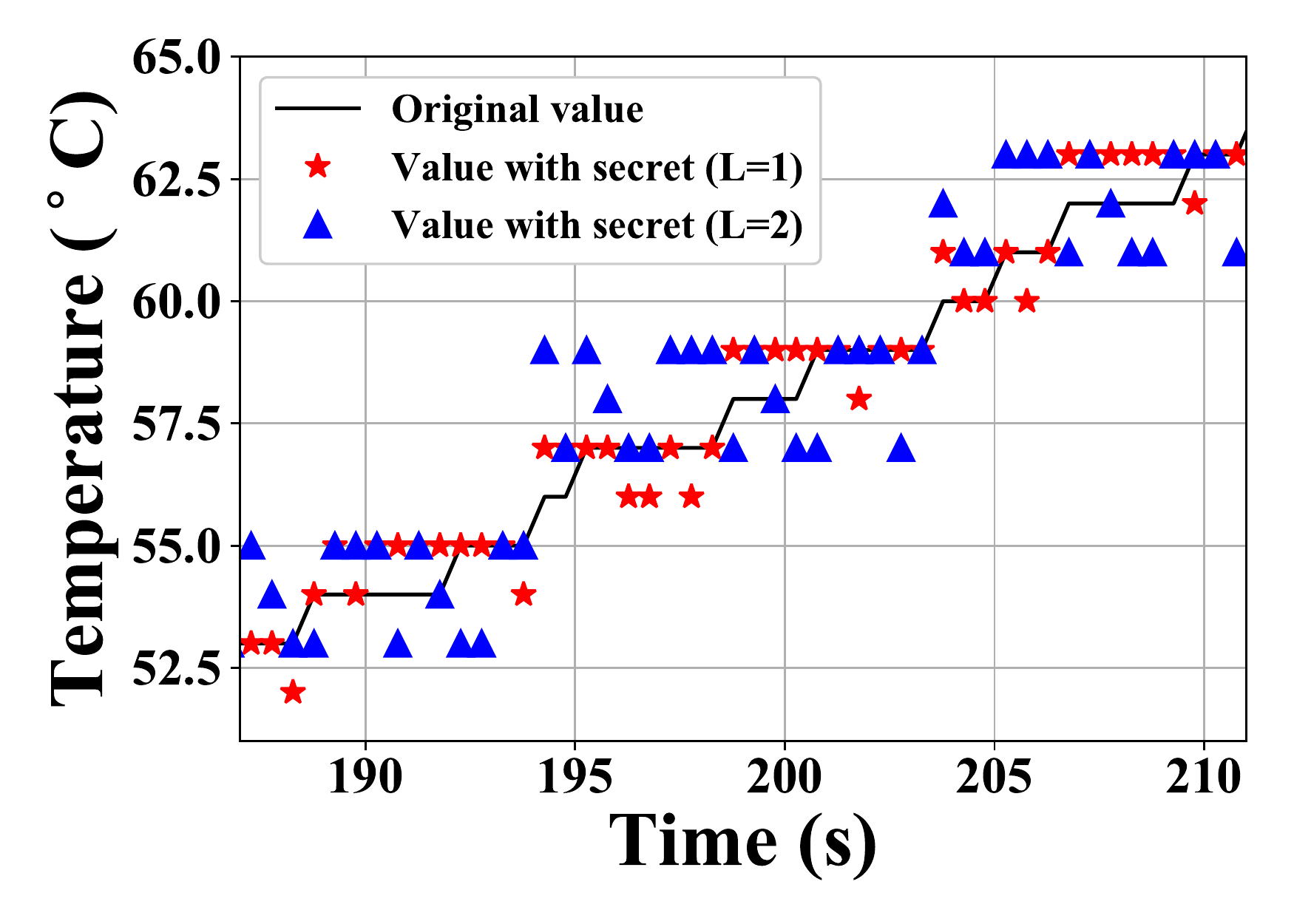}
    \caption{Authentication message embedded in EcoCAR engine coolant temperature data with $L=1$ and with $L=2$.}
    \label{fig:ecocar_lsb_evaluation}
    \vspace{-0.3cm}
\end{figure}

As illustrated in Figure~\ref{fig:toyota_lsb_evaluation} and highlighted in the magnified box, the maximum error introduced to wheel velocity is $0.01$ km/h for $L=1$ and $0.03$ km/h for $L = 2$, which is very insignificant. 
As for the engine coolant temperature as shown in  Figure~\ref{fig:ecocar_lsb_evaluation}, the maximum error is $1$ $^{\circ}C$ for $L = 1$ and $3$ $^{\circ}C$ for $L = 2$, which are still moderate.

\section{Conclusion and Future Work}
\label{sec:conclusion}
In this paper, we introduced TACAN, a transmitter authentication scheme using covert channels for the CAN bus, that allows a MN to verify the authenticity of the transmitting ECU. 
We developed IAT-based, offset-based, and LSB-based covert channels to communicate the authentication information between ECUs and the MN without introducing protocol modifications or traffic overheads. 
We provided a security discussion for TACAN and demonstrated the proposed covert channels through testbed validation and experimental evaluation.
Our future work will include experimental evaluation of TACAN under various attacks against the CAN bus.
In addition, we will also improve the throughput of the proposed covert channels and explore hybrid covert channel schemes.






\section{Acknowledgement}
\rev{
We thank the EcoCAR 3 project student lead Aman V. Kalia for helping us with the EcoCAR testbed. 
We also thank the anonymous reviewers for their helpful comments.}
This work was supported by NSF grant CNS-1446866, ONR grants N00014-16-1-2710 and N00014-17-1-2946, and ARO grant W911NF-16-1-0485. 
Views and conclusions expressed are that of the authors and not be interpreted as that of the NSF, ONR or ARO.

\bibliographystyle{ACM-Reference-Format.bst}
\bibliography{./references}

\newpage
\appendix
\section{Appendix}


\subsection{CAN Frame}
\label{appendix:CAN_frame}
\begin{figure}[ht!]
\centering
\includegraphics[width=1\columnwidth]{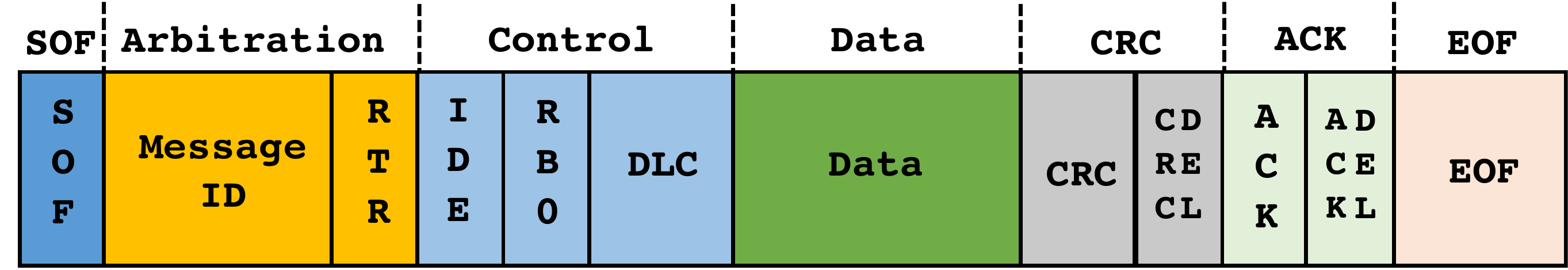}
\caption{Illustration of CAN frame structure.}
\label{fig:can_frame_structure}
\vspace{-0.3cm}
\end{figure}

As illustrated in Figure~\ref{fig:can_frame_structure}, each CAN frame or message has a set of predefined fields, including the Start of Frame (SOF) field, the Arbitration field (including a 11-bit message ID for the base frame format or a 29-bit message ID for the extended frame format), the Control field, the Data field (8-64 bits), the CRC field, the ACK field, and the End of Frame (EOF) field.

\rev{
\subsection{Impact on CAN Bus Schedulability}
\label{appendix:schedulability}
In order to understand the impact of TACAN on CAN bus schedulability, we apply the schedulability analysis in \cite{davis2007controller}.
For the ease of discussion, we provide a summary of definitions as below. 
The same notations in \cite{davis2007controller} are used for consistency.
\begin{itemize}
    \item $m$: Message priority (used interchangeably with message ID).
    \item $C_m$: Transmission time, defined as the longest time that the message can take to be transmitted.
    \item $J_m$: Queuing jitter, defined as the longest time between the initiating event and the message being queued, ready to be transmitted on the bus.
    \item $w_m$: Queuing delay, defined as the longest time that the message can remain in the CAN controller slot or device driver queue, before commencing successful transmission on the bus.
    \item $T_m$: Message period, defined as the minimum inter-arrival time of the event that triggers queueing of the message. 
    Such events may occur strictly periodically with a period of $T_m$ or sporadically with a minimum separation of $T_m$. 
    \item $R_m$: Worst-case response time of message $m$, defined as the longest time from the initiating event occurring to the message being received by the nodes that require it.
    \item $D_m$: Hard deadline, defined as the maximum permitted time from occurrence of the initiating event to the end of successful transmission of the message. 
\end{itemize}
A message is \textit{schedulable} if and only if its worst-case response time is no greater than its deadline ($R_m \leq D_m$). 
A CAN bus is schedulable if and only if all  messages on the CAN bus are schedulable.

As per \cite{davis2007controller}, we have $C_m = (80 + 10 s_m)\tau_{bit}$ (including bit stuffing), where $s_m$ is the number of data bytes and $\tau_{bit}$ is the transmission time of a single bit. 
For a $8$-byte message on a $500$ kbps CAN bus, we have $\tau_{bit}=2~\mu$s and  $C_m=320$ $\mu$s.
While CAN nodes typically have separate clock sources, all the timing quantities (e.g., message jitters, bit times, message periods, and deadlines) that derived from node clocks will be converted to real-time. 

The message $m$'s worst-case response time $R_m$ is given by
\begin{equation}
    R_m = J_m + w_m + C_m,
\end{equation}
and the queuing delay $w_m$ consists of two elements:
\begin{itemize}
    \item \textit{Blocking} $B_m$, due to lower priority messages  being transmitted when message m is queued, and
    \item \textit{Interference} due to higher priority messages which may win arbitration and be transmitted in preference to message $m$.
\end{itemize}
The blocking delay $B_m$ is given by $B_m = \max_{k \in lp(m)} (C_k)$, where $lp(m)$ is the set of messages with lower priority than $m$.

In order to analyze the worst-case response times, it is important to characterize the \textit{busy period}, in which all messages of priority $m$ or higher, queued strictly before the end of busy period, are transmitted during the busy period. 
The \textit{maximal} busy period begins with a so-called \textit{critical instant} where message $m$ is queued simultaneously with all higher priority messages and then each of these higher priority messages is subsequently queued again after the shortest possible time interval.

For simplicity, we assume that only one instance of message $m$ is transmitted during a priority level-$m$ busy period. 
In this case, the worst-case queuing delay is given by:
\begin{equation}\label{eq:worst_case_queuing_delay}
    w_m = B_m + \sum_{\forall k \in hp(m)} \left\lceil \frac{w_m + J_k + \tau_{bit}}{T_k}  \right\rceil C_k.
\end{equation}
Since the right hand side is a monotonic non-decreasing function of $w_m$, Eq.~(\ref{eq:worst_case_queuing_delay}) can be solved using the following recurrence relation,
\begin{equation}\label{eq:worst_case_queuing_delay2}
    w_m^{n+1} = B_m + \sum_{\forall k \in hp(m)} \left\lceil \frac{w_m^n + J_k + \tau_{bit}}{T_k} \right\rceil C_k.
\end{equation}
A suitable starting value is $w_m^0 = B_m$, and the recurrence relation iterates until, either $J_m + w_m^{n+1} + C_m > D_m$, i.e., the message is not schedulable, or $w_m^{n+1}=w_m^n$, in which case the worst-case response of message $m$ is given by $J_m + w_m^{n+1}+C_m$. 

In order to apply the above schedulability analysis to TACAN, we define $T'_m = T_m - \delta$. 
Since TACAN adds at most $\delta$ to each ITT, which can be considered as part of the queuing delay, we have $J'_m=J_m + \delta$. 
Assume that $\delta = 0.02 T_m$ and all messages employ the timing-based covert channel. 
By substituting $T'_m=0.98 T_m$ and $J'_m=J_m + 0.02 T_m$ into Eq.~(\ref{eq:worst_case_queuing_delay2}), we have
\begin{align}
    w_m^{' n+1} &= B_m + \sum_{\forall k \in hp(m)} \left\lceil \frac{w_m^{' n} + (J_k + 0.02 T_k) + \tau_{bit}}{0.98 T_k} \right\rceil C_k \nonumber\\
    &\approx B'_m  + 
    \sum_{\forall k \in hp(m)} \left\lceil \frac{w_m^{' n} + J_k + \tau_{bit}}{T_k} \right\rceil C'_k, \label{eq:updated_worst_case_queuing_delay}
\end{align}
where 
\begin{equation}
    B_m' = B_m + \sum_{\forall k \in hp(m)} \left(\frac{0.02}{0.98} C_k \right), 
\end{equation}
and 
\begin{equation}
    C'_k = \frac{1}{0.98} C_k = 1.02 C_k.
\end{equation}

Therefore, the effect of TACAN on schedulability is equivalent to increasing the blocking delay by a constant amount of time and increasing the  message transmission time by a certain percentage.
For example, if we assume $45$ messages with higher priority (half messages in the EcoCAR), the increase in the blocking delay and message transmission time is $294$ $\mu$s and $6.5$ $\mu$s ($C_m=320~\mu$s), respectively.  
By solving Eq.~(\ref{eq:updated_worst_case_queuing_delay}), we can compute the corresponding worst-case response time of message $m$ with TACAN. 
\rev{Hence, to achieve effective use of covert channels, TACAN parameters need to be experimentally obtained and fine tuned prior to deployment to ensure the schedulability of the CAN bus.}
}

\rev{
\subsection{LSB-Based Covert Channel Algorithms}
\label{appendix:lsb_algorithm}

In this section, we describe the authentication message embedding and extraction algorithms for the LSB-Based covert channel methodology. The related notations are summarized in Table~\ref{table:LSB_notation}.

\begin{table}[h]
	\footnotesize
	\centering
	\caption{Notation of LSB-based covert channel algorithms.}
	\begin{tabular}{|c|l|}
		\hline
		\textbf{Notation} & \textbf{Description} \\
		\hline
		$\alpha$  & Start sequence \\ \hline
		$A_m$ & Authentication (secret) message \\ \hline
		$A_f$ & Authentication frame \\ \hline
		$d_{field}(\cdot)$ & Data field of CAN data frame \\ \hline
		$\gamma$ & Selected bytes (value) \\ \hline
		$bin(\cdot)_L$ & Least significant $L$ bits \\ \hline
	\end{tabular}
	\label{table:LSB_notation}
	\normalsize
\end{table}

The embedding procedure is described in Algorithm \ref{pseudo:ecu_tx}.
For each $A_m$, a known preamble (a particular bit sequence) is first appended to the beginning of $A_m$ that allows the receiver to determine the start of $A_m$.
\rev{We assume that the authentication messages are generated and transmitted continuously.
Hence, the start of the next $A_m$ indicates the end of the previous one.}
Whenever new data field content $d_{field}$ is received from the upper (Application) layer, the least significant $L$ bits of selected values in $d_{field}$ (denoted as $bin(d_{field}(\gamma))_L$) are compared with the next $L$ bits of $A_f$: if the bits of interest are different then the substitution occurs, otherwise $d_{field}$ is not modified.
The same process is then repeated for each new authentication sequence. 

\SetAlgoNoLine
\begin{algorithm}[th!]
	\SetKwInOut{Input}{Input}
	\SetKwInOut{Output}{Output}
	\Input{$A_m$, $L$}
	$A_f \leftarrow \alpha||A_m$; $i \leftarrow 0$\;
	\While{$i < length(A_f)$}
	{
	    Receive $d_{field}$ from the upper (Application) layer\;
	    \If{$bin(d_{field}(\gamma))_L \neq A_f[i:i+L-1]$}
		{
		$bin(d_{field}(\gamma))_L = A_f[i:i+L-1]$\;
		}
		$i \leftarrow i + L$\;
		Send $d_{field}$ to the lower (Data Link) layer\;
	}

	\textbf{return}\;
\caption{Embedding $A_m$ to LSBs}
\label{pseudo:ecu_tx}
\end{algorithm}
\setlength{\textfloatsep}{3pt}

The extracting procedure is described in Algorithm \ref{pseudo:mon_rx}. 
When the MN receives a targeted CAN message, it filters for the Data frame and extracts from the $d_{field}$ the $L$ LSBs. 
Once a valid start sequence is detected, the MN starts storing the bits in $\hat{A}_f$ and comparing the newly arrived bits against the expected $A_f$.
In this way, it is possible to check possible inconsistency of the authentication sequence received also before the complete reception of $A_f$ to shorten the verification time.
If the actual value stored in $\hat{A}_f$ is not equal to part of the bits in $A_f$, an alert is raised highlighting a possible compromise of specific ECU or malicious exploitation of the CAN bus.
Once the $A_f$ is entirely received, the MN starts to extract the next authentication message.

\SetAlgoNoLine
\begin{algorithm}[ht!]
	\SetKwInOut{Input}{Input}
	\Input{$A_m$, $L$}
	$A_f \leftarrow \alpha||A_m$;  $i \leftarrow 0$\;
	\While{$i < length(A_f)$}
	{
	    Receive $d_{field}$ from the Data Layer layer\;
	    // Process $d_{field}$ in the Application layer\;
	    $\hat{A}_f[i:i+L-1]$ = $bin(d_{field}(\gamma))_L$\;
	    \If{$\hat{A}_f[i:i+L-1] \neq A_f[i:i+L-1]$}
		    {
		    Alert = True\;
		    }
		$i \leftarrow i + L$\;
	}
	\textbf{return}\;
	\caption{Extracting $A_m$ from LSBs}
	\label{pseudo:mon_rx}
\end{algorithm}
\setlength{\textfloatsep}{5pt}

}

\end{document}